\documentclass[sigconf,nonacm]{acmart}
\setcopyright{none}

\usepackage{booktabs}
\usepackage{tabularx}
\usepackage{array}

\usepackage[most]{tcolorbox}
\definecolor{featurebg}{HTML}{EAF1FB}   
\definecolor{featurefg}{HTML}{1A4E8A}   
\newcommand{\feature}[1]{%
  \tcbox[
    on line,            
    boxsep=1.2pt, left=2pt, right=2pt, top=0pt, bottom=0pt,
    arc=3pt,            
    colback=featurebg, colframe=featurebg,
    coltext=featurefg,
    fontupper=\sffamily\bfseries\small,
    nobeforeafter, tcbox raise base
  ]{#1}%
}

\usepackage{tikz}

\definecolor{verifiedbg}{HTML}{10B981}
\definecolor{verifiedframe}{HTML}{0D9F6E}

\definecolor{unverifiedbg}{HTML}{E4B555}
\definecolor{unverifiedframe}{HTML}{B28527}

\definecolor{greydotbg}{HTML}{D1D5DB}

\newcommand{\statusdot}[2]{%
  \tikz[baseline=-0.55ex]\draw[draw=#2, fill=#1, line width=0.4pt] (0,0) circle (0.45ex);%
}

\newcommand{\verifieddot}{\statusdot{verifiedbg}{verifiedframe}}
\newcommand{\unverifieddot}{\statusdot{unverifiedbg}{unverifiedframe}}
\newcommand{\halfverifieddot}{%
  \tikz[baseline=-0.55ex]{%
    \begin{scope}
      \clip (0,0) circle (0.45ex);
      \fill[verifiedbg] (-0.45ex,-0.45ex) rectangle (0,0.45ex);
      \fill[greydotbg] (0,-0.45ex) rectangle (0.45ex,0.45ex);
    \end{scope}
    \draw[draw=verifiedframe, line width=0.4pt] (0,0) circle (0.45ex);
    \draw[draw=verifiedframe, line width=0.3pt] (0,-0.45ex) -- (0,0.45ex);
  }%
}

\usepackage{booktabs}
\usepackage{tabularx}
\begin{document}

\title{LeanSide: A Formally Verified Co-Reasoning System for Natural-language Proofs}

\newcommand{\berkeleyaffiliation}{%
  \affiliation{%
    \institution{University of California, Berkeley}
    \city{Berkeley}
      \state{CA}
      \country{USA}
  }%
}

\author{Chenjun Guo}
\berkeleyaffiliation
\email{chenjun\_guo@berkeley.edu}

\author{Manooshree Patel}
\berkeleyaffiliation
\email{manooshreepatel@berkeley.edu}

\author{Arnav Mehta}
\berkeleyaffiliation
\email{arnavmehta@berkeley.edu}

\author{Krishiv Kothari}
\berkeleyaffiliation
\email{krishivkothari@berkeley.edu}

\author{Thomas Lu}
\authornote{These authors contributed equally to this work.}
\berkeleyaffiliation
\email{thomaslu@berkeley.edu}

\author{Niels Voss}
\authornotemark[1]
\berkeleyaffiliation
\email{niels\_voss@berkeley.edu}

\author{Rayna Bhattacharyya}
\authornotemark[1]
\berkeleyaffiliation
\email{rayna\_b@berkeley.edu}

\author{Peter Donovan}
\authornotemark[1]
\berkeleyaffiliation
\email{peterdonovan@berkeley.edu}

\author{Bjoern Hartmann}
\berkeleyaffiliation
\email{bjoern@eecs.berkeley.edu}

\author{Gireeja Ranade}
\berkeleyaffiliation
\email{ranade@eecs.berkeley.edu}

\renewcommand{\shortauthors}{Guo et al.}

\begin{abstract}
Large language models are increasingly used as collaborators on deductive-reasoning tasks, but their outputs can hallucinate or pull users away from intended reasoning. Formal proof assistants provide machine-checked verification, but have a steep learning curve and require more granular reasoning than human written proofs. We explore an interface that combines these strengths, allowing users to write and revise free-form natural-language proofs while a verified backend checks their reasoning and returns feedback at the user's granularity. We study this interface in the context of undergraduate mathematics education by developing LeanSide, a formally verified co-reasoning system, which auto-formalizes student reasoning into Lean and informalizes verifier output into understandable feedback. We conducted user studies through classroom deployment and analyzed which system properties helped students make progress and which caused them to get stuck. We use these findings to derive design implications for using a formally verified backend in human-AI co-reasoning systems.
\end{abstract}





\maketitle

\section{Introduction}
Large language models (LLMs) are being explored as collaborators on deductive-reasoning tasks, where conclusions necessarily follow from given premises or rules~\cite{tarski1936logical}. Such tasks arise across domains, including mathematics~\citep{yang2024formal,liu2025safeenhancingmathematicalreasoning,ospanov2025hermes,hendrycks2021measuring,cobbe2021training,wei2022chain,lewkowycz2022solving}, programming~\citep{hendrycks2021measuring,li2022competition}, legal judgment~\citep{guha2023legalbench,jurayj2026languagemodelslogicprograms,pramaanalabs2026}, and clinical decision support~\citep{singhal2023large,kung2023performance,nori2023capabilities,singh2026guarantees}.

However, LLM-generated reasoning can contain hallucinations~\citep{xu2023large,ji2023survey}. One solution is to express reasoning in a machine-checkable formal language such as Lean~\citep{deMouraUllrich2021lean4}, Isabelle~\citep{paulson1994isabelle}, and Rocq~\citep{barras1997coq}, whose kernels mechanically checks formalized proofs and provides hallucination-free feedback. Such formal representations are widely used in high-assurance domains, including hardware verification~\citep{witharana2022survey}, and increasingly in the formalization of mathematical research~\citep{paulson2023large,shulman2024strange}. A successfully checked formal artifact provides strong confidence in logical validity.

But these languages have a steep learning curve~\citep{shi2025qedobsstudy,thoma2022lean} and demand far more granular and rigorous reasoning than people conventionally write~\citep{patel2026leantutorverifiedaimathematical,guo2025unspoken,tarrach2024more}. A growing body of work has started to use auto-formalization to automatically formalize informal reasoning into machine-checkable code, originating from and richest in mathematics~\citep{ying2024lean,azerbayev2023proofnet,wu2022autoformalization,lu2024process,liu2025rethinking,patel2026leantutorverifiedaimathematical}, and extending to other rule-based domains, such as legal and tax judgment~\citep{jurayj2026languagemodelslogicprograms,pramaanalabs2026} and clinical support~\citep{singh2026guarantees}.



Despite rapid progress in auto-formalization, prior systems have primarily aimed to generate verified formal code for increasingly advanced problems, rather than to faithfully follow a user's reasoning trace. They have also paid less attention to interactive settings where user input may be free-form, ambiguous, incomplete, or still in progress. Traditionally, formalization is a post-hoc check: a mathematician formalizes a finished proof into machine-checkable code after the argument is complete~\citep{wu2022autoformalization,lu2024process}. Existing auto-formalization systems largely inherit this framing, typically only support well-formed, textbook-style proofs as input~\citep{lu2024process,liu2025rethinking,azerbayev2023proofnet}. As a result, they are not well suited to serving as interactive collaborators for proofs that are still being developed.

\begin{figure}[t]
    \centering
    \includegraphics[width=1 \linewidth]{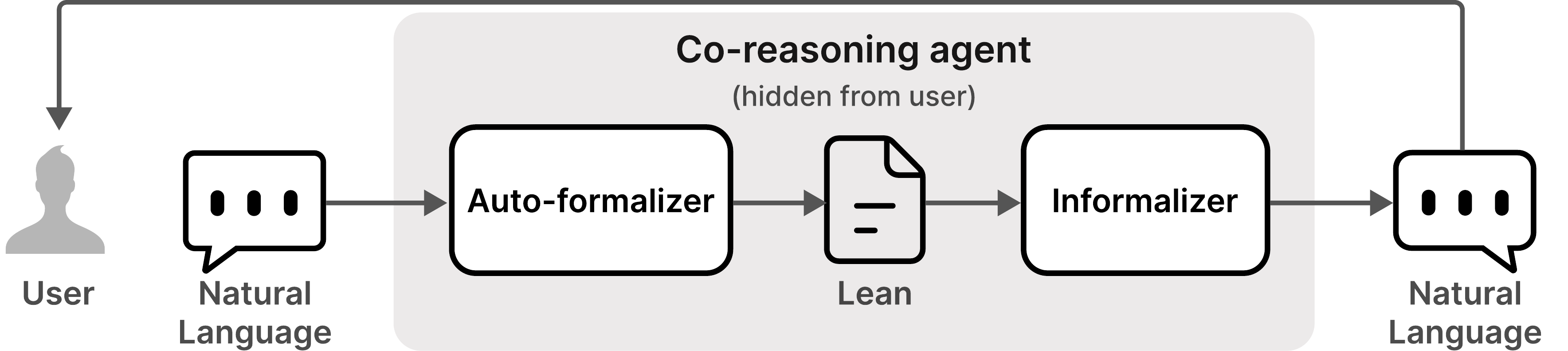}
    \vspace{-5pt}
    \caption{User workflow in a co-reasoning system: users express input and receive output in natural or informal language while the system auto-formalizes input into Lean, verifies it, and informalizes the result.}
    \label{fig:user_workflow}
\end{figure}

In human communication, partners continually build mutual understanding~\cite{clark1991grounding, clark1986referring}, thus a formally verified collaborative agent must support an analogous grounding loop: users need to know both whether the system has ``understood'' their reasoning and what verification result it produced. \textbf{We aim to build an informal-reasoning collaborative agent that makes
the power of formal verification accessible to broader audiences 
who are trained to
reason in their own domains but have no knowledge of the formal language.} 
Users read and write as they would when communicating with another person, while a verified backend interprets and checks their reasoning: auto-formalization maps informal input to a formal representation, and informalization maps verifier output back into everyday reasoning (Figure~\ref{fig:user_workflow}). This interaction allows users to gain confidence in the validity of the system's feedback.

We study student interactions with such a formally verified co-reasoning system (FVCR) for undergraduate mathematics, a timely setting as students increasingly use LLM-based tools for homework feedback~\citep{chen2026chat,wardat2023chatgpt, liu2023novice}. 
LLMs can produce convincing but at times incorrect answers~\citep{xu2023large,ji2023survey,kasneci2023chatgpt,wang2023examining} or distract students from their original reasoning paths~\citep{liu2023novice}.

We call this tool
\textbf{LeanSide}. \textbf{LeanSide affords students to express their reasoning in familiar language and at their desired granularity (i.e., the level of detail of a step),
validate their reasoning through formally verified feedback, and iterate on mistakes in their proofs}. Our goal is not to provide fully ``tutor-like'' feedback: we do not optimize the system to behave like a human tutor or to be maximally helpful pedagogically.
Instead, we investigate the possible role of formal verification in an interaction loop where students work using familiar informal language, asking how much Lean's verification power can help students. We also recognize that recent high-profile AI mathematics results do not make formal verification obsolete; they often pair generative search with formalization~\citep{openai2026navierstokes,deepmind2024alphaproof}, and our work asks how this pairing can be brought into interactive settings for students and other users. 

\begin{figure*}[t]
    \centering
    \includegraphics[width=0.6\linewidth]{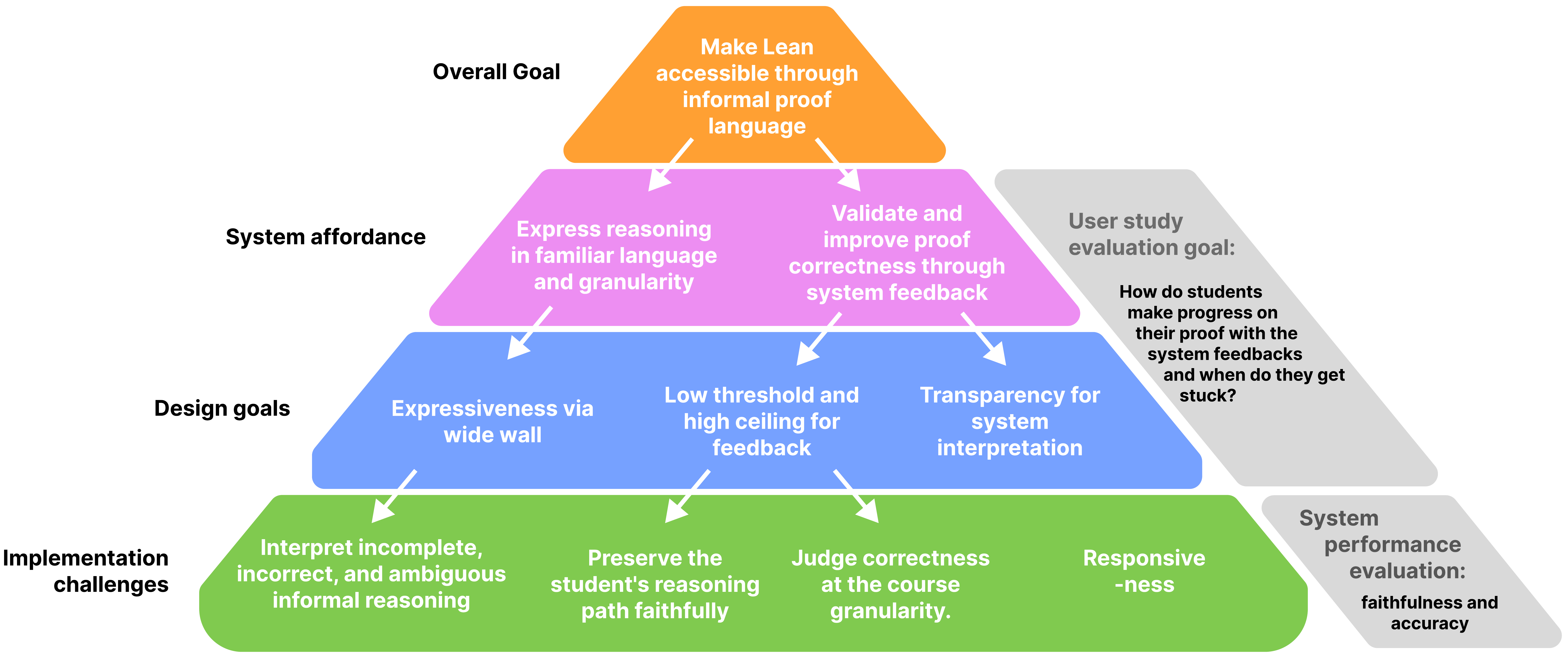}
    \caption{This figure summarizes the system's goals, affordances, design goals, and implementation requirements.}
    \label{fig:goals}
\end{figure*}

The main goals and affordances are shown in Fig.~\ref{fig:goals}. We contribute: (1) LeanSide, a system that uses formal language as a verification backbone while supporting the design goals of low threshold, high ceiling, wide walls, and transparency for non-expert interaction. We describe the technical challenges of meeting these goals and our implementation solutions. (2) A user study with 23 participants, conducted alongside a classroom deployment, examining whether and how students use system feedback to iterate on their proofs. (3) Empirical findings on which system features helped students make progress and which barriers limited progress, together with design implications for how AI co-reasoning systems can use formal verification to move beyond current limitations.

\section{Related Work}
In this section, we first discuss relevant HCI literature on co-reasoning systems. We then turn to recent advances and limitations in auto-formalization, which lays the background and motivation of our works but have largely been discussed outside HCI in the mathematics, AI and education communities.

\subsection{Challenges of human-AI collaboration in reasoning tasks}
 
The recent popularity of LLMs has renewed the aspiration of using natural language to collaborate with machines on reasoning tasks. LLM based system has long being discussed about the problems of hallucinations and user agency preservation.

In LLM-assisted reasoning, fluent explanations can make incorrect conclusions particularly difficult to reject~\citep{kim2025fostering}, users may become more confident after accepting incorrect advice~\citep{bo2025rely}, and erroneous LLM-generated information can directly lead to incorrect decisions~\citep{spatharioti2025effects}. 
In logical reasoning tasks, programmers~\citep{nguyen2024beginning} and domain scientists~\citep{o2025scientists} could both fail to detect errors in generated code. Prior interfaces address this problem by exposing inconsistencies~\citep{cheng2024relic}, highlighting suspected hallucinations~\citep{leiser2024hill}, or making generated behavior executable and inspectable~\citep{ferdowsi2024validating}. In our system, we take another route to reducing hallucinations by using a formally verified backend.

Prior work has also long raised concerns about preserving user agency in such collaboration. Although current interfaces appear to support fully free-form natural-language input, users, especially novices, can struggle to articulate their thoughts into effective informal reasoning input for the system~\citep{zamfirescu2023johnny,nguyen2024beginning,jiang2022discovering}. In human-AI collaboration for programming, users also face difficulty inspecting whether generated code interpreted and implements their intended reasoning~\citep{zi2025written} and debugging prompts systematically~\citep{zamfirescu2023johnny,jiang2022discovering,zi2025written}. Prior work links these challenges to failures of shared understanding: plausible but misaligned LLM outputs can pull users away from their original reasoning and require costly monitoring and repair~\citep{vila2025miscommunicating,subramonyam2024bridging,mozannar2024reading}. In programming, long LLM generated suggestions can disrupt programmers' flow~\citep{barke2023grounded}, while inspecting, verifying, and editing them imposes additional interaction costs~\citep{mozannar2024reading}. In decision-making, automated suggestions can reduce users' initiative~\citep{levy2021assessing}, and LLM recommendations can leave less room for users' independent reasoning~\citep{reicherts2025ai}. Across broader domains, LLM-based collaboration can also weaken users' sense of ownership~\citep{draxler2024ghostwriter} or redirect their thinking toward the model's suggestions~\citep{jakesch2023co}.

Consequently, interfaces have begun to externalize the system's interpretation to bring mutual understanding back to the conversation.
\citet{liu2023wants} frame natural-language programming as an abstraction-matching problem and help users inspect and repair mismatches between their goals and the system’s interpretation, closely paralleling our design. 
SOVITE~\citep{li2020multi} aims to build mutual understanding with the users by displaying the system’s inferred intents through GUI screenshots. These works motivate our focus on making the system's interpretation inspectable.

\subsection{Lean and auto-formalization}

Interactive theorem provers such as Lean~\citep{deMouraUllrich2021lean4}, Isabelle~\citep{paulson1994isabelle}, and Rocq~\citep{barras1997coq} provide machine-checked environments for formalizing mathematics. However, they remain difficult for non-experts: they require a steep learning curve and a level of granularity far below ordinary mathematical writing~\citep{shi2025qed}, creating an accessibility gap between formal verification and how students and mathematicians normally write proofs.

Auto-formalization aims to narrow this gap by translating natural-language mathematics into formal language. Recent work has focused mainly on statement auto-formalization, which translates informal theorems into Lean propositions~\citep{wu2022autoformalization,azerbayev2023proofnet,weng2025autoformalization,jiang2023multilingual,ying2024lean,gao2024herald,liu2025atlas,wu2026stepfun,lu2024process,li2024autoformalize,lu2025formalalign,liu2025rethinking,poiroux2025reliable}, and automated theorem proving, which starts from a formal proposition and generates compiled formal proofs~\citep{yang2023leandojo,xin2024deepseekprover,lin2025goedel}.

A smaller body of work explores whole-proof auto-formalization~\citep{jiang2022draft,huang2024mustard,murphy2024autoformalizing,wang2024theoremllama,tarrach2024more,patel2026leantutorverifiedaimathematical,cabral2025proofflow}, raising the issue of faithfulness: whether the formal proof preserves the content and organization of the informal reasoning. Existing work operationalizes faithfulness in several ways, including semantic similarity between informal and formal steps~\citep{li2024autoformalize,lu2025formalalign}, bidirectional equivalence between formal candidates~\citep{liu2025rethinking,poiroux2025reliable}, structural fidelity of intermediate claims and proof dependencies~\citep{patel2026leantutorverifiedaimathematical,cabral2025proofflow}, and robustness to paraphrases or invalid steps, where systems must avoid silently changing or repairing the user's meaning~\citep{moore2025evaluating,cornish2026faithformbench}. At the proof level, ProofFlow represents proof structure as a dependency graph~\citep{cabral2025proofflow}.

However, most auto-formalization benchmarks still assume polished, textbook-style, one-shot inputs. ProofNet pairs textbook problems with a single natural-language proof and Lean statement~\citep{azerbayev2023proofnet}; miniF2F and Lean Workbook focus on theorem or problem statements rather than human proof revision~\citep{zheng2021minif2f,ying2024lean}; and several widely used datasets are constructed from formal corpora rather than situated human proof writing, such as MMA and Herald~\citep{jiang2023multilingual,gao2024herald}, TheoremLlama's OBT data~\citep{wang2024theoremllama}, and MUSTARD~\citep{huang2024mustard}.

Student proofs in learning contexts are often more fragmented, nonlinear, implicit, and loosely structured than these benchmarks assume~\citep{wolska2004analysis,guo2026bridging}. Moreover, when students work with a human tutor or tutoring system, they often expect iteration, making their proofs more likely to be incomplete, incorrect, or expressed only as high-level ideas~\citep{vanlehn2006behavior,benzmueller2007dialog}. We therefore study auto-formalization not as one-shot translation of finished proofs, but as a co-reasoning process through which students introduce, revise, repair, and clarify mathematical ideas while the formal state evolves with the interaction.

\subsection{Making formal verification accessible}

Prior work has explored multiply ways to make formal verification more accessible to both expert and non-experts. In mathematics education, several studies have tried to make formal languages accessible by teaching them directly in mathematics classrooms~\citep{hanna2019proof,hendriks2010teaching,henz2011teaching,jacobsen2023exams,thoma2022learning}. Students benefit from faster feedback loops, but these tools still pose pedagogical challenges: Lean error messages can be insufficient for students~\citep{hanna2019proof}, and proof-assistant feedback may signal that a problem exists without explaining why~\citep{aldrich2008sasylf}. Hazel Prover addresses this challenge through design criteria for educational proof assistants, including pedagogical feedback, error tolerance, scaffolding, easy interaction, and classroom deployability~\citep{keenan2024hazel}. These criteria offer an educational perspective complementary to our focus.

Other systems lower the barrier through visualization and controlled natural language. These include visualizing unsolved Lean goals~\citep{nawrocki2023extensible}, de-automating automatically generated proofs~\citep{shi2025designing}, supporting controlled-natural-language proofs~\citep{wemmenhove2022waterproof, massot2024teaching}, and using domain-specific visualizations to teach formal methods~\citep{nelson2024forge}. These tools make formal systems easier to inspect and use, but still only allowed a subset of utterances.

A related line of work studies human-AI collaboration for proof formalization. Lea supports mathematician-in-the-loop Lean development~\citep{lea2026}, while LLMStep~\citep{welleck2023llmstep} and Lean Copilot~\citep{song2024lean} assist formalization with tactic suggestion, proof search, and premise selection. These systems primarily support experts producing formal proofs, rather than non-experts seeking verified feedback on natural-language reasoning.

Formal verification has also been applied beyond mathematics, including tax, legal~\citep{jurayj2026languagemodelslogicprograms, pramaanalabs2026}, and medical reasoning~\citep{hekmatnejad2019modelchecking, singh2026guarantees}. These works demonstrate the potential of utilizing machine-checkable reasoning in high-stakes domains, but primarily focus on the technical realization rather than learning how people engage with formally verified feedback through informal-language interfaces.

Our work is related to all of these efforts, but moves beyond expert-facing tools by using a natural-language interface and examining how people interact with such systems in practice.

\begin{figure*}[t]
    \centering
    \includegraphics[width=1\linewidth]{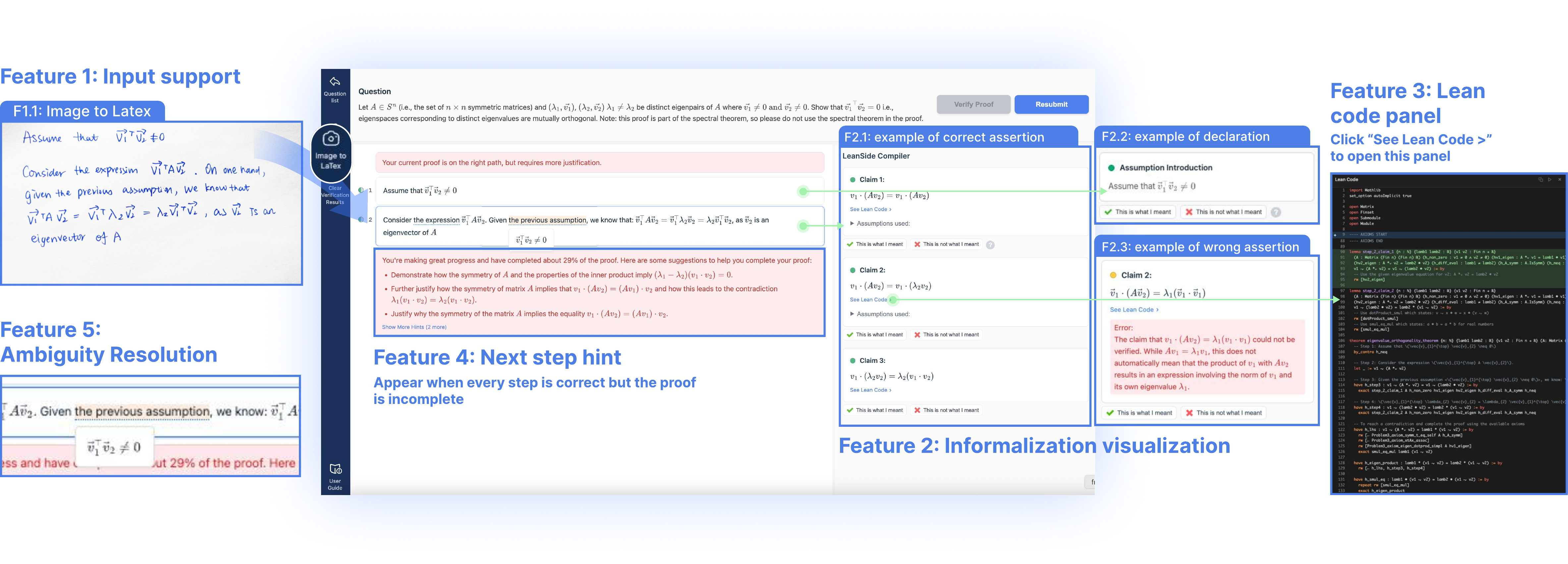}
    \vspace{-30pt}
    \caption{The User Interface. Students use the system by uploading handwritten proofs, or typing in. In this snapshot, a student has already uploaded an image of their proof and completed the verification process. The resulting feedback is displayed.}
    \label{fig:vision}
    \vspace{-10pt}
\end{figure*}

\section{System design}

This system targets undergraduate students with no knowledge of Lean, and it is designed for deployment in undergraduate mathematics courses. 
To afford expressing reasoning through natural language,
validating and iterating on proofs, we adopt
three corresponding design goals: (1) expressiveness via wide
walls~\cite{resnick2005design}, (2) low-threshold and high-ceiling
feedback~\cite{myers2000past}, and (3) transparency of system representation.

A user's workflow is as follows: (1) the user enters free-form reasoning by uploading handwriting, or typing directly; (2) The system auto-formalizes the input into verified Lean code (3) the resulting code and compiler output are informalized into natural-language feedback; and (4) the user revises their proof based on the feedback. Users can iterate on their proof without reading or writing Lean, but can open a Lean code panel to see how their steps map to Lean if interested.

\subsection{Expressiveness via wide walls}
Creative-support tools should have ``wide walls,'' supporting various forms of exploration beyond predefined actions~\cite{resnick2005design}. We treat proof-solving as a creative activity---students use different strategies, write at different granularity, and rely on various informal or ambiguous representations. Our system therefore supports reasoning in whichever form users find most natural. Users can upload a photo of handwritten work \feature{F1} (see Fig. \ref{fig:vision} for this and all feature refs) or type directly in \LaTeX{} on central panel. These methods covers two primary ways our target users write proofs. Unlike many auto-formalization settings, where natural-language proofs are textbook-like, complete, and close to formal structure, input modes in our system support proofs as they are produced, not only after they are polished, even when they are ambiguous, exploratory, incorrect, or incomplete.

\subsection{Low Threshold and High Ceiling}

Effective tools should be easy for novices to start with while supporting advanced work~\cite{myers2000past}. Our system aims to make Lean's verification power accessible without requiring Lean expertise. 

First, the interface helps users connect their written reasoning to the system's formalization: clicking a user-written proof-step cell in the central panel displays the corresponding informalized Lean interpretation in the right panel \feature{F2}. A natural-language step may map to multiple informalization cells when it contains multiple smallest progression units (further explained in Section \ref{pre-processing}), as in \feature{F2.1} where step 2 maps to two assertions. 
Each smallest progression unit is classified as either a declaration or an assertion. Declarations, such as assumption introductions \feature{F2.2}, are always marked green \verifieddot{} because users may introduce assumptions freely. Assertions are marked green \verifieddot{} when verified, or yellow \unverifieddot{} with an error message when verification fails \feature{F2.3}. 
The informalization visualization keeps feedback at \textit{the user's granularity}: specific enough to localize errors, but not so fine-grained that low-level formal details create information overload.

Second, the interface makes errors understandable without requiring users to inspect Lean code. Step-wise errors are shown directly inside the informalization cell for the incorrect claim \feature{F2.3}; when all written steps are correct but the proof remains incomplete, the system shows a continuation feedback at the end of the user's proof \feature{F4}. 

Advanced users who know Lean or want to learn it can click ``See Lean code'' to view the highlighted Lean block correspond to each step \feature{F3}. 

Here is an overview of five possible feedback types. Appendix \ref{appendix_more_examples} shows examples for each type of feedback.

\begin{itemize}
    \item \textbf{Step-level error messages} appear when some steps are faulty; each unverified step shows an informalized proposition and an error message.
    \item \textbf{Continuation error messages} appear when all steps are verified but the proof remains incomplete, under the conditions described in Section~\ref{prover}.
    \item \textbf{Continuation hints} also appear when all steps are verified but the proof remains incomplete, under the conditions described in Section~\ref{prover}.
    \item \textbf{Large-jump warnings} provide optional guidance on verified steps that could be further justified.
    \item \textbf{Suggestions} provide optional guidance for fully verified, correct, and complete proofs, helping make the proof more rigorous.
\end{itemize}


\subsection{Transparency}
\label{system_design_transparency}
Free-form input alone is insufficient: the system must make its interpretation transparent. Users can trust verification feedback only after confirming that the system formalized the reasoning they intended. 
We do this at two levels.
First, users are able to assess the faithfulness of auto-formalization through Feature \feature{F2}, which maps each user-written proof step to its informalized Lean interpretation. Second, to support free-form input, the system surfaces how it resolves underspecified language: ambiguous references are highlighted with predicted meanings \feature{F5}, and each sentence is labeled with its predicted action type, such as an assertion or declaration \feature{F2.2}. We describe the pre-processing module behind these predictions in Section~\ref{pre-processing}.

\section{Implementation}

To support expressiveness, validation, and proof iteration, the system must meet
four implementation challenges:

\begin{itemize}
    \item \textbf{Interpret free-form reasoning.} The system must be able to
    interpret student reasoning even when it is incorrect, ambiguous, or
    incomplete.

    \item \textbf{Preserve the student's reasoning path faithfully.} The system must
    faithfully auto-formalize and informalize student reasoning. This differs
    from common auto-formalization settings, where the primary goal is to
    produce compiled Lean code for a target theorem. In our system, silently
    repairing a flawed argument or altering the proof path would undermine the
    student's ability to see whether their own reasoning was understood and
    verified.

    \item \textbf{Judge accurately at the course granularity.} The system must
    accurately determine whether each proof step is valid. In an educational
    setting, correctness also depends on granularity: a proof may be formally
    valid but pedagogically unacceptable if it skips an instructor-defined
    intermediate step. The system therefore needs to identify both incorrect
    steps and steps whose reasoning gap is too large for the course context.

    \item \textbf{Responsiveness.} The system must return feedback quickly enough to support an interactive co-reasoning loop. Commercial auto-formalization systems that handle research-level mathematics can be slow \cite{lau2026aristotlecase}, while proof tutoring requires students to revise, resubmit, and compare feedback repeatedly. The system therefore balances verification quality with latency suitable for real-time interaction.
\end{itemize}

\subsection{System pipeline}

\begin{figure*}[t]
    \centering
    \includegraphics[width=1\linewidth]{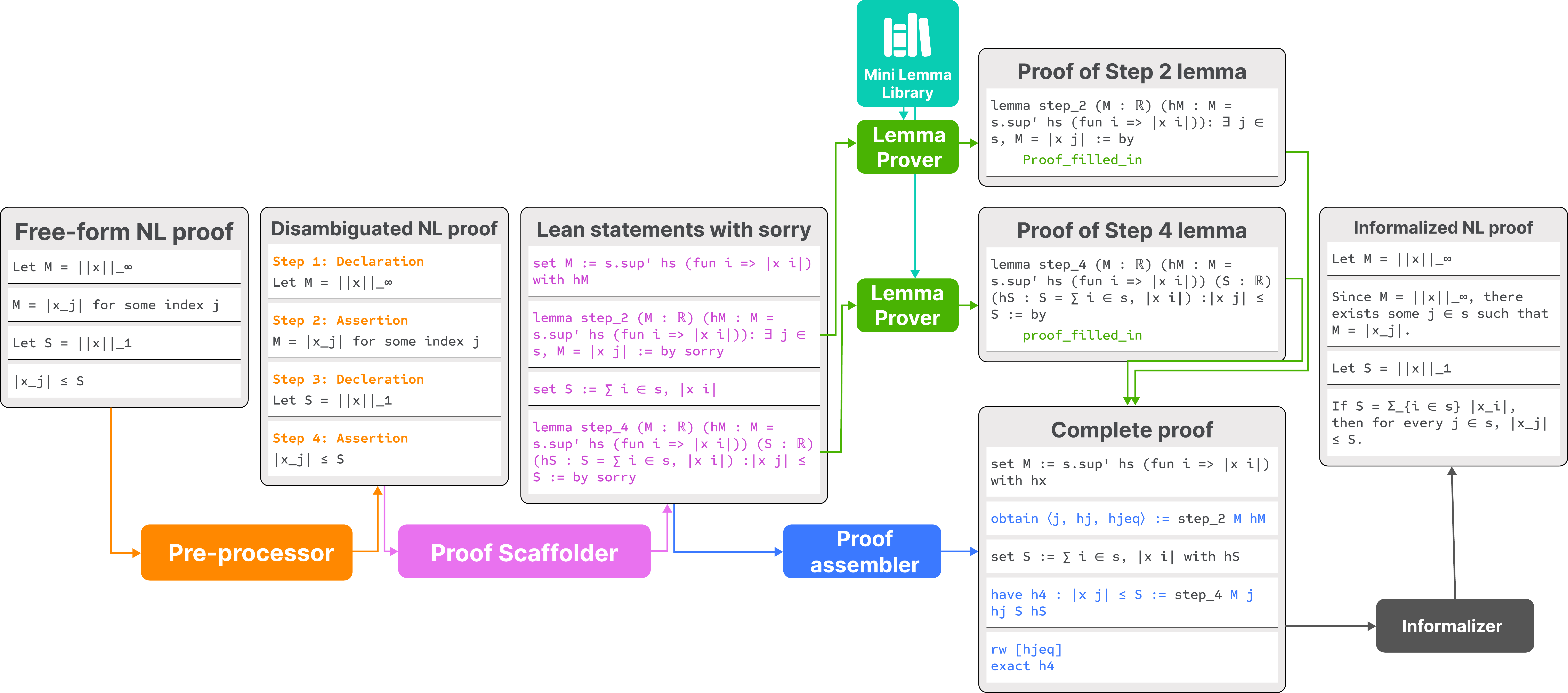}
    \caption{This figure shows the system pipeline. Student input (left) is auto-formalized into Lean through the pre-processor, proof scaffolder and prover, which includes the lemma prover and proof assembler. The generated Lean code and error messages are then informalized back into natural language. The mini-lemma library supports the lemma prover.}
    \label{fig:system_pipeline}
\end{figure*}

The system consists of two main components: an auto-formalizer, which translates students' proof attempts into Lean and attempts to compile them, and an informalizer, which translates the generated Lean code and compiler feedback back into natural language for students. The auto-formalizer includes three modules: a natural-language pre-processor, a proof scaffolder, and a prover. All the modules are LLM-based. Figure \ref{fig:system_pipeline} explains the detailed architecture of the system.

\subsubsection{Natural-language pre-processor.} \label{pre-processing}
The natural-language pre-processor is the first module in the auto-formalizer. It takes students' original natural-language input, either typed directly into the system or entered through Optical Character Recognition (OCR). This module builds on the pre-processing system of Guo et al.~\cite{guo2026bridging}. It has two goals: (1) to resolve ambiguity in the natural-language proof so that the proof can be more reliably translated into Lean, and (2) to parse the proof into the smallest progression units. We define a \textit{smallest progression unit} as a minimal, self-contained statement, at the level of content expressed by the student, that includes one antecedent, one consequent, and evidence(s). 

Our system resolves three common ambiguities in student proofs through a three-stage chain adapted from Guo et al.~\cite{guo2026bridging}:

\begin{itemize}
    \item \textit{Referential ambiguity.} The system first identifies referentially ambiguous terms whose intended referents are unclear and predicts those referents. At this stage, the proof remains a single piece of text and is not yet parsed into steps.

    \item \textit{Action-type ambiguity.} The system then parses the proof according to action type, where the ambiguity is whether a step functions as an assertion or a declaration. Assertions progress the proof by deriving new information from known results, hypotheses, or unsolved goals, while declarations set up necessary context without directly progressing the proof~\cite{guo2026bridging, sieg1990proof}. Based on student proofs in practice, we further divide declarations into five subtypes: introducing new variables, stating the proof goal, declaring a proof strategy, repeating known facts from the question or earlier proof, and declaring completion by restating the final conclusion. The proof is segmented whenever the action type changes.

    \item \textit{Antecedent-consequent-evidence ambiguity.} Finally, the system parses continuous assertions into the smallest progression units and predicts the antecedent, consequent, and evidence for each unit. This step addresses ambiguities where students write multiple assertions continuously and leave some of these components implicit.
\end{itemize}

To support transparency, the interface displays the results of these ambiguity-resolution steps as described in Section \ref{system_design_transparency}. 
\subsubsection{Proof Scaffolder}
\label{proof_scaffolder}
This module takes the pre-processor output and converts each smallest progression unit into a self-contained Lean 4 lemma statement mapped one-to-one to the student's proof. It does this through declaration scaffolding and assertion scaffolding. Declaration scaffolding converts informative declarations, such as introduced variables and stated proof goals, into Lean code, while ignoring declarations that repeat known facts or restate the final conclusion. Assertion scaffolding converts each assertion into a Lean proposition, represented as \verb|prop| in the lemma template \verb|lemma name (hypotheses): prop := by sorry|. Following the proof order, each later assertion's \verb|(hypotheses)| includes all hypotheses from the problem and from the aforementioned declarations. The scaffolder does not prove these propositions; it fills the proof body with \verb|sorry|, a placeholder that lets Lean provisionally accept the proposition.

This design separates proposition generation from lemma proving to better preserve faithfulness to the student's proof structure. 

\subsubsection{Prover}
\label{prover}
The prover receives Lean declarations and assertion propositions in the order expressed by the student. It consists of two components: a lemma prover, which attempts to prove each assertion lemma, and a proof assembler, which combines the verified lemmas to check whether they complete the proof. If any assertion fails in the lemma prover, the proof assembler is not triggered; instead, the error message is passed to the informalizer for student-facing feedback. Both provers are LLM-based and use an iterative loop: when generated code fails to compile, the prover retries using the code and compiler error message until the proposition compiles or the maximum number of attempts is reached.

\paragraph{Mini-lemma library}
\label{mini_lemma_library}
To satisfy the system requirement of responsiveness, we use faster LLMs as the prover backbone. However, such models usually have weaker reasoning ability, we complement them with a curated mini-lemma library. The library contains two types of lemmas: rubric lemmas, which are mapped to the rubric for each problem, and side-support lemmas, which capture smaller pieces of reasoning that students are not expected to write in the course context but that the prover cannot reliably generate.

We curated the mini-lemma library semi-automatically. Researchers first used stronger LLM models to generate Lean proofs for rubric lemmas, then tested the library on full-credit student proofs from previous exams, and added side-support lemmas when the prover failed on those proofs. This process helped the library support multiple student proof strategies. We also developed a fully automated lemma-generation pipeline to improve generalizability, although human inspection is still required to label rubric and side lemmas. The mini-lemmas for Question~13 (Appendix \ref{appendix:problems}) were curated using this fully automated pipeline, with two researchers reviewing the lemma quality to ensure the same standard as for the other questions.
\subsubsection{Informalizer}

The informalizer generates student-facing feedback from the lemma propositions and compiler error messages. Its outputs are shown in the informalization cells (\feature{F2}) and feedback panel (\feature{F4}). For each step, the informalizer translates the Lean proposition back into natural language using researcher-curated instructions design to aligned with the notation and terminology used in the course. Note that the informalizer takes only the Lean code and error message as input, but do not see the student's proofs. We specifically designed the system this way to use Lean's verification ability without relying on the LLM's pedagogical ability.

For step-level feedback, a successfully verified step is marked green \verifieddot. If the prover used more than one rubric lemma, the system informalizes those lemmas and shows a large-jump warning. If the step cannot be verified, the Lean compiler error is informalized into a step-wise error message. 

For whole-proof feedback, the system considers two cases. If the proof assembler fails to compile the problem statement using the student’s steps, the proof is marked incomplete and the compiler error is informalized into a continuation error message. If the proof assembler succeeds, the system counts the rubric lemmas used in the assembled proof. When more than two rubric lemmas are used, the proof is still treated as incomplete and the system shows a continuation hint. When two or fewer rubric lemmas are used, the proof is treated as complete, and the system may show a suggestion based on the supporting and rubric lemmas used.

\section{User study procedure}
This section describes the study context and procedure.

\subsection{Context: classroom deployment}

We deployed the system in an advanced undergraduate linear algebra course at a university, where most students were computer science majors, with a smaller portion from other engineering majors or graduate programs.
Across 13 weeks, we deployed 14 questions, releasing one to five questions weekly or biweekly according to the course curriculum. Before deployment, we conducted three pilot studies with course teaching assistants or students who had previously taken the course.

At the start of the semester, research team gave a 10-minute in-class demonstration of how to use the system. As new questions were released, we conducted one-on-one user study sessions with students. Students participated by voluntarily signing up for open slots. Participants received 1 point of extra credit, out of 100 total course points, for attending the session; their performance and whether they completed the proof during the session did not affect this credit. The system logged all student actions. 

\subsection{Participants}
In total, we conducted 27 user study sessions. Two sessions were unusable for technical reasons, one for recording failure and one for a student who was not following the class context and thus was not able to understand and attempt to answer the question.
Therefore, the analyses presented in this paper are based on the remaining 23 sessions. 
Participants had a similar distribution of majors (14/23 are Computer Science), class years (21/23 are undergraduate students), and grades as the full class (details in Appendix \ref{appendix_participants}).

22 participants had used the system at least once before attending the study session, while the remaining one knew how to use it from the classroom demonstration. 6 participants reported that they had heard of Lean, and 2 reported that they had tried Lean through an introductory game~\cite{buzzard2023naturalnumbergame}, but none of the participants had sufficient Lean knowledge to write a proof, even one simpler than the course material, on their own. 

\subsection{One-on-one User study protocol}
Each study session lasted two hours. We first spent 5 minutes establishing rapport and asking participants about their backgrounds and experience. Then, during the first 1.5 hours, students wrote proofs and iterated on them using the system's feedback. In the remaining half hour, we conducted a short interview about their experience with the system.

In the first 1.5 hours, students were prompted to choose questions to complete during the study. They were told that there was no time limit for any individual question and that the study included no reward or penalty based on whether they answered questions correctly. If a participant finished one question, they were asked to start another until the allotted time was exhausted. We used three criteria for selecting questions: (1) the student must not have previously attempted the question using the system; (2) we encouraged students to choose questions for which they were less certain about the solution, so that we could observe iterative progress; and (3) when both criteria were satisfied, we preferred more recently released questions that were closer to the material just taught in class. No participant completed all questions correctly on the first attempt without seeing system feedback.

Participants were allowed to consult course materials, their own notes, and browser search results, but could not use AI tools. When search engines displayed AI-generated overviews, we allowed participants to read them as part of the search results, provided they had not entered the proof question directly as a query.

Participants were instructed to think aloud while interacting with the system. However, they were not required to think aloud while writing proofs, to avoid disrupting their proof construction. Instead, each time participants finished a proof and before they submitted it for verification, we asked them to briefly walk through their reasoning and rate their confidence in the proof's correctness on a 1--5 scale.
Each time participants received system feedback, we asked three questions: (1) whether each informalization accurately represented what they meant (faithfulness); (2) whether the system feedback was accurate; and (3) for error messages, whether they were understandable, relevant, and useful. Participants were prompted to comment on any of these topics and were also free to raise other comments.

In the final half hour, participants were interviewed about their experience using the system, including how it compared with using an LLM or attending an in-person office hour. Participants who had not previously opened the Lean code panel were also invited to open it once. The researchers briefly explained how to read the code, and participants were asked whether the code provided any useful extra information not visible in the platform.

\begin{figure*}[t]
    \centering
    \includegraphics[width=1\linewidth]{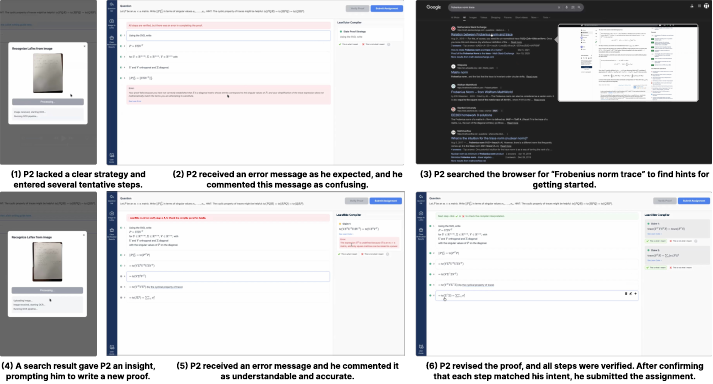}
    \caption{An usage example of how student iterates on one problem from start to end}
    \label{fig:usage_example}
\end{figure*}

\subsection{Analysis and Evaluation}

We extracted five types of information from each user study session: (1) expert-evaluated system performance: whether the system's feedback was faithful and accurate according to the definitions in the implementation section; (2) perceived system performance: whether the student thought the system had auto-formalized faithfully and provided accurate feedback; (3) proof progress: whether each student modification was effective, incorrect, or unrelated; (4) students' open-ended comments after receiving each piece of feedback; and (5) students' answers to the predefined interview questions.

Assessing (1) and (3) required mathematical knowledge of the course material. We therefore recruited two researchers who had taken or audited the course and were familiar with the course material. The two researchers independently judged 20\% of the sessions (6 participants). When disagreements arose, we consult the course instructor to make the final decision. We refer to these two researchers and the instructor as the expert group. For (2), we used think-aloud data; when students did not comment, we prompted them with questions such as ``Do you think this claim is actually what you meant?'' and ``Do you think this error message is correct?''. We then made a binary judgment for each informalization. For (4) and (5), we followed a reflexive thematic analysis approach, beginning with inductive coding of a subset of sessions, iteratively refining candidate codes and themes, and then using the resulting themes to guide analysis of the remaining data.

\section{An example of the system usage}

To illustrate the complete interaction trajectory, we present an example of one participant's end-to-end interaction on a problem that includes both helpful and unhelpful feedback. We also encourage the reader to watch the attached video figure for better understanding about the user workflow. Participant 2 (P2) began without a clear strategy and entered several tentative steps, including repeated hypotheses and familiar facts. After clicking \textit{Verify}, P2 received an error message that he found
confusing and unhelpful for making progress (reasons in Section \ref{system_performance_understandbility}), so he searched the browser for guidance. He identified a plausible strategy, rewrote the proof on paper, and uploaded it again. In subsequent iterations, P2 judged the feedback as accurate, corrected a step accordingly, and eventually verified every step. A top banner then prompted him to check whether each system interpretation matched his intent; after checking every step, he submitted the assignment.

\begin{figure*}[t]
    \centering
    \includegraphics[width=1\linewidth]{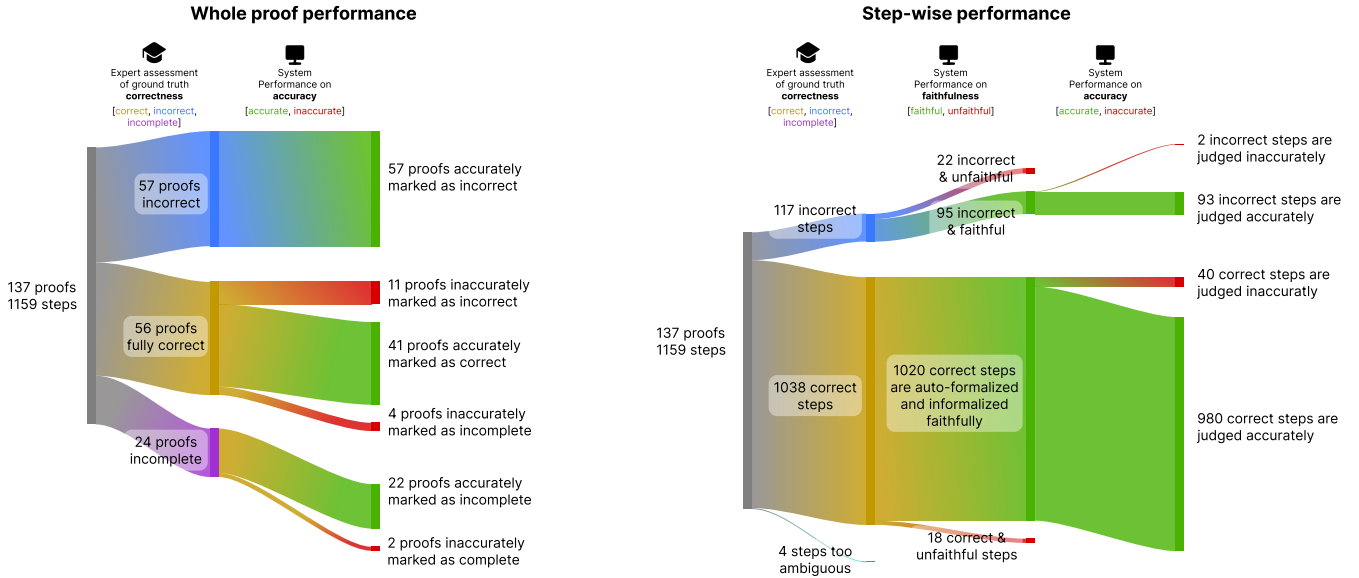}
    \caption{This figure shows the system's faithfulness and accuracy performance: the left side shows whole-proof performance, and the right side shows step-level performance. 
    We use \textit{correct} and \textit{incorrect} for expert-assessed ground truth, \textit{accurate} and \textit{inaccurate} for whether the system's judgment matches that ground truth, and \textit{faithful} and \textit{unfaithful} for whether auto-formalization preserves the student's original input.
    }
    \label{fig:system_performance}
\end{figure*}

\section{Prerequisites for natural-language access to verification}

We listed faithfulness and accuracy as the requirement for Lean to be accessible through natural language, thus, in this section, we evaluate the system's performance on these two criteria. The expert group manually reviewed all natural-language proofs produced in the user studies and the generated Lean code from all study sessions. This section reports the system's performance. Figure~\ref{fig:system_performance} shows the whole proof performance (left), and the step-wise performance (right). In this figure and the following sections, we use \textit{correct} and \textit{incorrect} to denote the expert assessment of a student step's ground truth correctness; \textit{accurate} and \textit{inaccurate} to denote whether the system's judgment matches that ground truth; and \textit{faithful} and \textit{unfaithful} to denote whether the auto-formalization preserves the student's original input.

For whole-proof performance (Figure~\ref{fig:system_performance}, left), we collected 137 proofs: 56 (41\%) were fully correct, 57 (42\%) contained at least one erroneous step, and 24 (18\%) had only correct steps but were incomplete. The system maintained a very low false-positive rate for detecting incorrect and incomplete proofs, while marking 11/56 (20\%) fully correct proofs (according to the rubric) as containing an error and 4/56 (7\%)fully correct proofs as incomplete.

As the faithfulness and accuracy is measured by steps, we also calculate the total number of steps\footnote{the steps used here is measured by minimal independent step defined in the Section \ref{pre-processing}, as the student divided step do not have a fix size, some student put all reasoning in one cell, while some put each one equation in one cell, so to avoid inconsistency, the system performance is measured using the minimal independent step.} demonstrated in Figure ~\ref{fig:system_performance}, right. In total, serving as ground truth,  we have 1038 expert evaluated correct steps, 117 (11\%) wrong steps and 4 steps which are too ambiguous to judge correctness. The correct number is more than incorrect ones as in an incorrect proof, there is often only one or a few steps are wrong, while the rest steps are valid. The following paragraphs evaluate step-level performance and explain the reasons for unfaithfulness and inaccuracy.

\subsection{System performance on faithfulness} \label{system_performance_on_faithfulness}

Expert group judged that 18/1038 (2\%) of correct steps and 22/117 (18\%) of incorrect steps are unfaithful. Most unfaithful cases arose during auto-formalization. Three cases involved errors in both auto-formalization and informalization, and one edge case due to informalization unfaithfulness alone.

Auto-formalization unfaithfulness had two main causes. One common cause (12 cases) occurred when a student made a small error in an incorrect step, such as omitting a transpose and therefore multiplying matrices with incompatible dimensions. When the proof scaffolder module (Section~\ref{proof_scaffolder}) tried to auto-formalize the step, the Lean statement could not compile unless this error was fixed, causing the scaffolder's compiler-guided loop automatically
corrected it. Despite the scaffolder prompt's instruction to preserve the student's meaning even when it is incorrect, these small formal invalidities were the pipeline's most likely source of unfaithful auto-correction. Section \ref{Perceive_unfaithful} explains how does student judge whether the auto-formalization of one step is unfaithful and Section \ref{start_with_incorrect_proof} further explains what does the student do when facing unfaithful auto-formalized steps.

The second cause occurred when students wrote steps that were vague or ambiguous along one of the three ambiguity types, mathematically trivial, or not
intrinsically propositional, and the auto-formalization produced a stronger claim that human evaluators judged inconsistent with the student's intended meaning. Section \ref{start_with_ambiguous_proof} explains how student make use of these stronger claims to progress on their proofs.

\subsection{System performance on accuracy} 
\label{system_performance_on_accuracy}
\subsubsection{Step-wise accuracy} Measuring accuracy for unfaithful steps is not reasonable, so the accuracy is only measured within the faithful steps. Expert judge that for 40/1020 (4\%) of correct steps, the system is being extra stingy or not accurate and judged correct step as wrong. The expert group judged step-level accuracy with respect to each
question's rubric: if a student step was more granular than a rubric item, the step was judged incorrect even when it was mathematically valid.

Inaccuracy had two main causes. First, the prover (Section~\ref{prover}) sometimes failed to prove a mathematically valid proposition at the expected
granularity, either because the step followed an uncommon proof strategy that had not been tested when constructing the mini-lemma library, or because the step was mathematically valid but not relevant to any solution strategy. In both
cases, because the mini-lemma library was built around expected solution paths, it did not contain the lemmas needed to prove the step. Since our prover prioritizes responsiveness over proving arbitrary mathematical propositions, these steps failed to verify. 

Second, students sometimes did not explicitly declare the types or dimensions of new variables. For example, in a problem involving singular value decomposition, a student might write
$P = U \Sigma V$ without specifying the types and dimensions of $U$, $\Sigma$,
and $V$; the proof scaffolder could then formalize them as arbitrary variables,
such as real numbers. Although humans often omit these details as unspoken
conventions, the system may fail under this under-specification. This case could be classified as either unfaithful or inaccurate; here, we label it inaccurate.
For both causes, we expect the inaccuracy rate to decrease as the underlying
technology improves. Section \ref{inaccurate} further explains how the student judge whether the system accurately judged their step.

In theory, the system would never judge an faithfully auto-formalized incorrect step as correct, as lean would never able to compile it, however, in real usage, there is one edge case which judged incorrect step as correct, because that step has been wrongly classified as an assumption---repeat known, as mentioned in Section \ref{proof_scaffolder}, the repeat known assumptions are not formally verified.

\subsubsection{Error-message understandability}
\label{system_performance_understandbility}
Even when the system accurately judged a step's correctness, its error message did not always explain the mistake in an understandable way. The expert group therefore separately evaluated whether each error message identified the step's mistake in the way a human collaborator would. To isolate the usefulness of the Lean error itself from the pedagogical value of an LLM, the informalizer prompt
included only the Lean error message and Lean lemma declaration, not the student's original proof.

We evaluated the understandability of each feedback types. Expert judges found that 61/97 (63\%) of step-level error messages accurately reflected what was wrong in the step, compared with 3/13 (23\%) of continuation error messages and (11/13)85\% of continuation next-step hints.

Step-level error messages were difficult to understand mainly because of a granularity mismatch. When students wrote a statement as a single assertion, they often treated it as not needing further explanation in the course context.
Lean errors, however, often reflected failures in lower-level proof obligations, making the feedback more granular than the student's step, thus not understandable. 

Continuation error messages had a similar limitation: when the proof assembler failed to complete the Lean code for the proof, it returned a reason why the whole proof was not closed, but this reason was not necessarily at the user's granularity. By contrast, continuation next-step hints were more understandable because they were informalized from the mini-lemma library, which was mapped to rubric items and designed to stay within the course context.

\section{User study findings}
\begin{figure*}[t]
    \centering
    \includegraphics[width=0.8\linewidth]{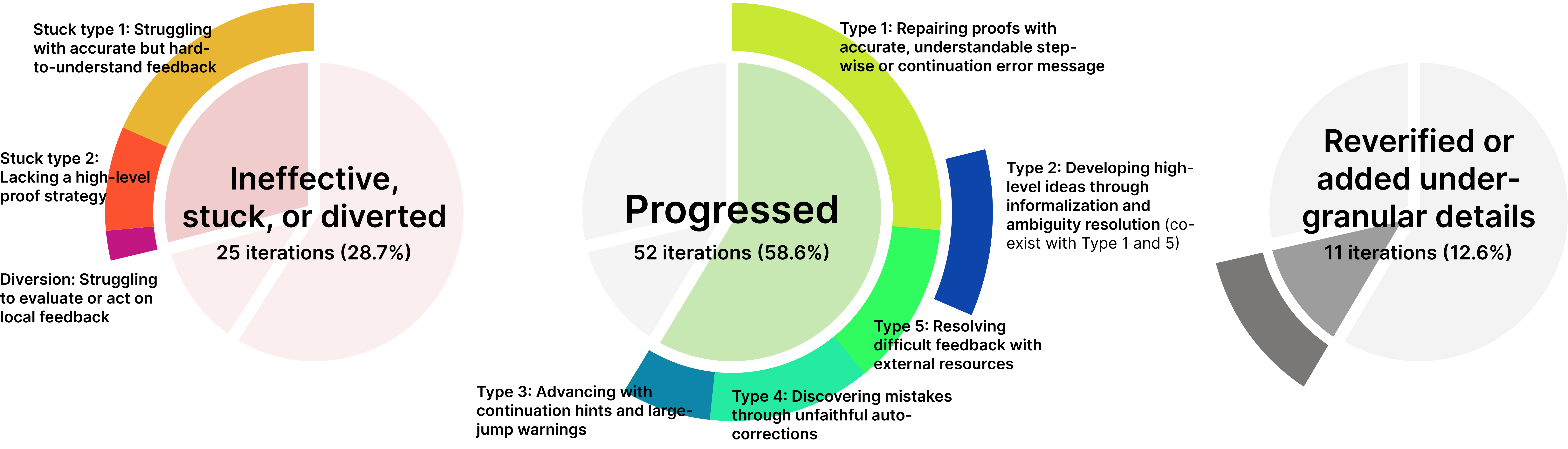}
    \caption{Distribution of coded revision outcomes. Most iterations led to progress (52 iterations, 58.6\%), while 25 iterations (28.7\%) were not useful, involved getting stuck, or diverted students from productive proof paths, and 11 iterations (12.6\%) involved reverifying or adding under-granular details. The labels show the progress and breakdown types analyzed in this section.}
    \label{fig:result_data}
\end{figure*}

This section examines how students unable to author Lean proofs independently accessed Lean-backed verification through LeanSide. We report on (1) how students assessed the system’s faithfulness and accuracy, (2) when they progressed or got stuck, and (3) when the interaction increased their confidence.

In each iteration, students fell into one of three outcomes: they made progress, became stuck or diverted, or, when facing possibly inaccurate feedback, they re-verified the same proof or added more details to help the system understand their proof. Students submitted 137 proofs across 50 problems. After excluding final submissions that did not lead to another iteration, we coded 87 iterations (Figure~\ref{fig:result_data}): 52 (58\%) led to effective progress, 25 (29\%) were not useful or involved getting stuck or diverted, and 11 (13\%) involved reverification or adding more granular details to support the prover.


\subsection{Could participants assess the system's accuracy and faithfulness?}

To revise their proofs using system feedback, participants first needed to judge whether the feedback faithfully represented their reasoning and was mathematically accurate. We compared their judgments with those of our expert group. Agreement was high: participants matched the experts on 87\% of faithful steps and 83\% of unfaithful steps, as well as 91\% of accurate feedback and 99\% of inaccurate feedback.

\subsubsection{How do participants judge a claim as inaccurate?}
\label{inaccurate}
We observed two main ways students identified inaccurate step-level or continuation error messages. Most commonly, students relied on their own mathematical judgment, especially when the disputed step was easy to verify from the local context --- for example, when it directly applied a theorem taught in class. Lean could nevertheless fail to verify such steps when the required theorem was absent from Mathlib or our mini lemma library (Section~\ref{mini_lemma_library}), or when a trivial informal claim required a non-trivial formal proof. Thus, steps that students considered easy were not always easy for Lean to verify. P2, P11, P19 all encountered steps that they were confident were correct despite receiving error messages. P11 explained this local judgment process:
\begin{quote}
\leftskip=1.5em
\rightskip=1.5em
\itshape
I might not be confident that the whole proof is correct, but it's much easier to decide if one step is correct. When I first look at the error, I will try to diagnose whether the error is on my side or on the system's side. The error message is local to the claim, so if I can confirm the claim is correct, then I will assume it's a system error.
\end{quote}

Second, students recognized an error message as inaccurate when it clearly fell outside the local mathematical context or operated at a finer granularity than their intended proof. One example is the inaccurate continuation error message discussed in Section~\ref{stuck_lack_high_level} and shown in Figure~\ref{fig:eg4}.

Without either cue, students had less basis for evaluating the feedback and were more likely to pursue an incorrect direction. When students judged an error message as inaccurate (11 iterations), they either (1) re-verified the same input in the hope that the prover would verify it, (2) added more detail to help the prover, or (3) stopped once they had localized the issue to a step they were confident was correct.

\subsubsection{How do participants judge a claim is unfaithful?} \label{Perceive_unfaithful}
Students primarily judged faithfulness by assessing whether their original idea and the informalization were mathematically equivalent or could be trivially converted into each other. In practice, this often began with noticing that the informalization “looked different” from their original input. However, informalizations rarely looked exactly the same as the input, even when they were faithful. 
Thus, judging faithfulness required students to use their mathematical knowledge to decide whether the input and output were equivalent.
We observed three cases in which students did not notice that a claim was unfaithful, all of which led to unproductive iterations. This usually occurred when the system proved an unfaithful claim and marked the step as correct, causing the indicator beside the step to appear green \halfverifieddot{}. Students typically paid more attention to gray steps with error messages and revised those steps first, as a result, they sometimes missed the unfaithful claims. In Section \ref{discussion_faithfulness}, we discuss the design challenge of prompting users to check faithfulness, as well as the design decisions we explored and their effects on users’ attention to faithfulness.

\subsection{How do participants make progress?}
\subsubsection{Progress type 1: Repairing proofs with accurate, understandable step-wise or continuation error message (23 iterations)}

\begin{figure}[t]
    \centering
    \includegraphics[width=1\linewidth]{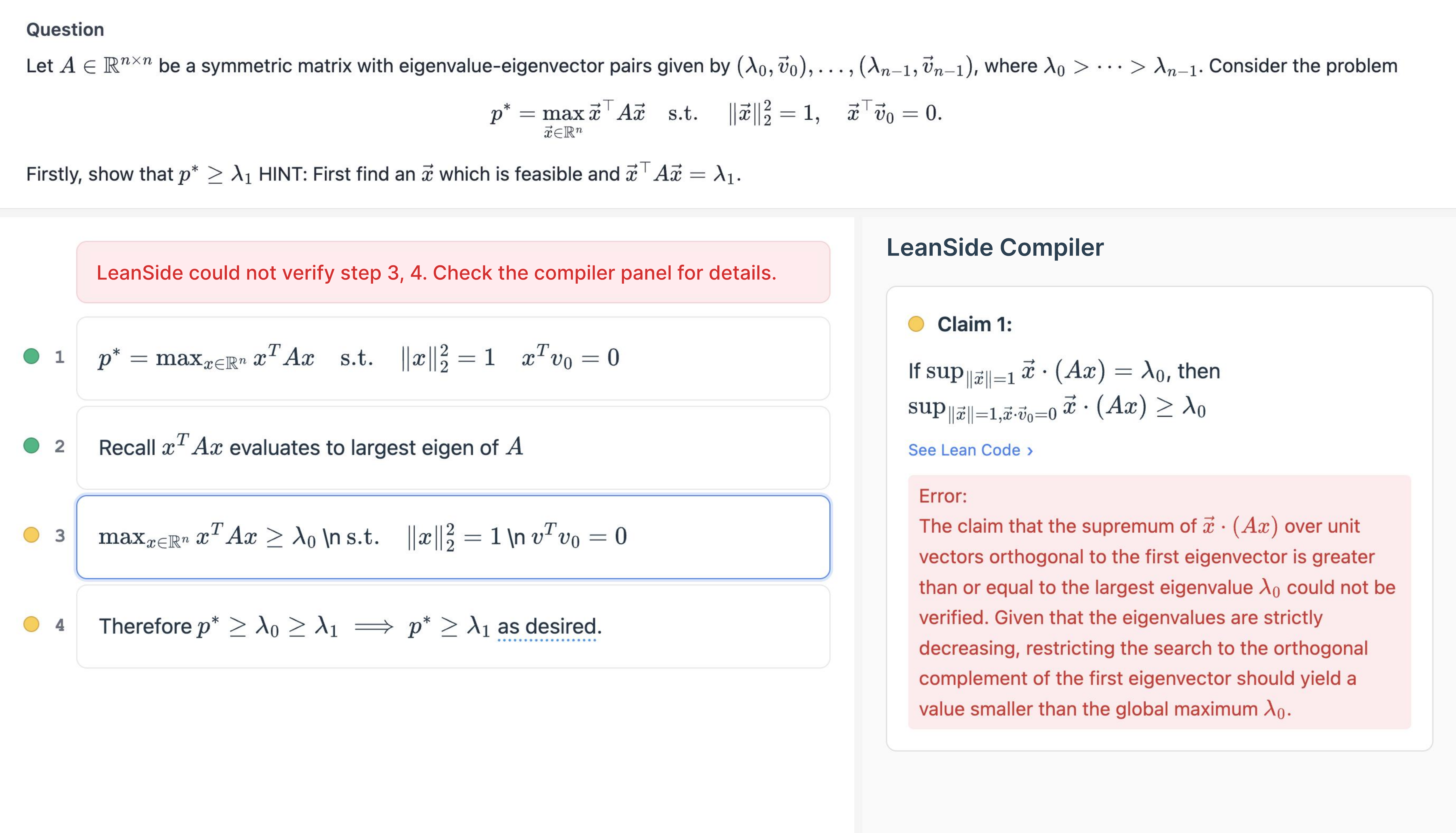}
    \vspace{-5pt}
    \caption{This figure shows an understandable and accurate error message output by the system.}
    \label{fig:eg1}
    \vspace{-15pt}
\end{figure}
A common starting point was a proof containing error(s) the student had not recognized. Step-level error messages, when accurate and understandable, helped students repair both major misconceptions and subtle mistakes. For Q13, P12 overlooked a key constraint (Figure~\ref{fig:eg1}. The error message drew attention to this omission: \textit{``Oh right! Because $x^\top$ is also supposed to be orthogonal to the first eigenvector, it actually can't be $\lambda_0$.''} P12 then revised the proof strategy. The system also caught subtle errors, of which we include an example in Appendix \ref{appendix_more_examples}.

Understandable continuation error messages could similarly help students complete unfinished proofs. 
These messages were generally effective when students had already made valid progress and needed only one or two additional rubric items. For Q8, P2 recognized that the Cauchy--Schwarz inequality was relevant but did not know how to apply it. The continuation error message explained its role, allowing P2 to close the logical gap and continue the proof (More details about this example is explained in Appendix \ref{appendix_more_examples}).

\subsubsection{Progress type 2: Making high-level ideas more rigorous through informalization and ambiguity resolution (9 iterations)}
\label{start_with_ambiguous_proof}
Students also began with a conceptual intuition but were unsure how to express it mathematically. After first reading a problem, P27 explained, \textit{It's hard for me to translate an expression before I have fully formulated my thinking.''} Students therefore submitted descriptive or tentative reasoning and used the system's interpretation to develop it. P14 described a proof as \textit{fairly conceptual,''} adding, \textit{Even in the exam, maybe the grader would not accept that, but maybe a mathematician will say this looks good. Maybe demonstrate this part more clearly.''} Similarly, P12 wrote, \textit{This proof is hand-wavy and I yap a lot. Let's see what the system says and we could iterate.''} The system helped students turn such ambiguous expressions into rigorous arguments in two ways.

First, when an expression matched one of the three ambiguity types described in Section~\ref{pre-processing}, the interface displayed the system's predicted interpretation for students to inspect. Even incorrect predictions were useful because they revealed where an expression had confused the system. For example, in several cases (P2, P22, P24, and P26), action labels in the informalization cells helped students assess whether their intent had been captured. P26 initially wondered, \textit{I am not sure whether the system could understand I'm repeating the definition, or should I delete these steps?''} Seeing the ``Repeat Known'' label confirmed that the system had understood the step. 
In two cases, antecedent, consequent and evidence ambiguity resolution also helped (P19, P4). For example, P19 used ``\textit{Hence}'' twice in one step, but the first instance referred to earlier reasoning rather than the immediately preceding step. The generated proposition did not match his intent, revealing the ambiguous antecedent. P19 therefore noticed and resolved this ambiguity.

Second, informalization helped students develop conceptual ideas that fell outside these predefined ambiguity types. For Q13 (Appendix \ref{appendix:problems}), P18 input an ambiguous, high level proof idea into the system with some missed cases:
\begin{quote}
\leftskip=1.5em
\rightskip=1.5em
\itshape
Since the quadratic form
\(y^{T}Ay = \sum_{i=0}^{n-1}\lambda_i (y\cdot e_i)^2,\)
we know that the weight assigned to each eigenvalue is the dot product between $y$ and the corresponding eigenvector. When $y=v_1$, this means all the weight is assigned to $\lambda_1$. In other scenarios, the weight is distributed across the other eigenvectors. However, in these other cases, it is strictly better to allocate more weight to the largest eigenvector since doing so yields the largest weighted sum.
\end{quote} 
These steps were not mathematically precise statements, but were along a path that could get to a rigorous proof. The informalization solidified the student intuition through propositions propositions representing the weight allocation process and normalization constraint on the weights, and suggested a rigorous approach to the proof.
P18 commented that this \textit{``makes the proof more robust.''}
\begin{figure}[t]
    \centering
    \includegraphics[width=1\linewidth]{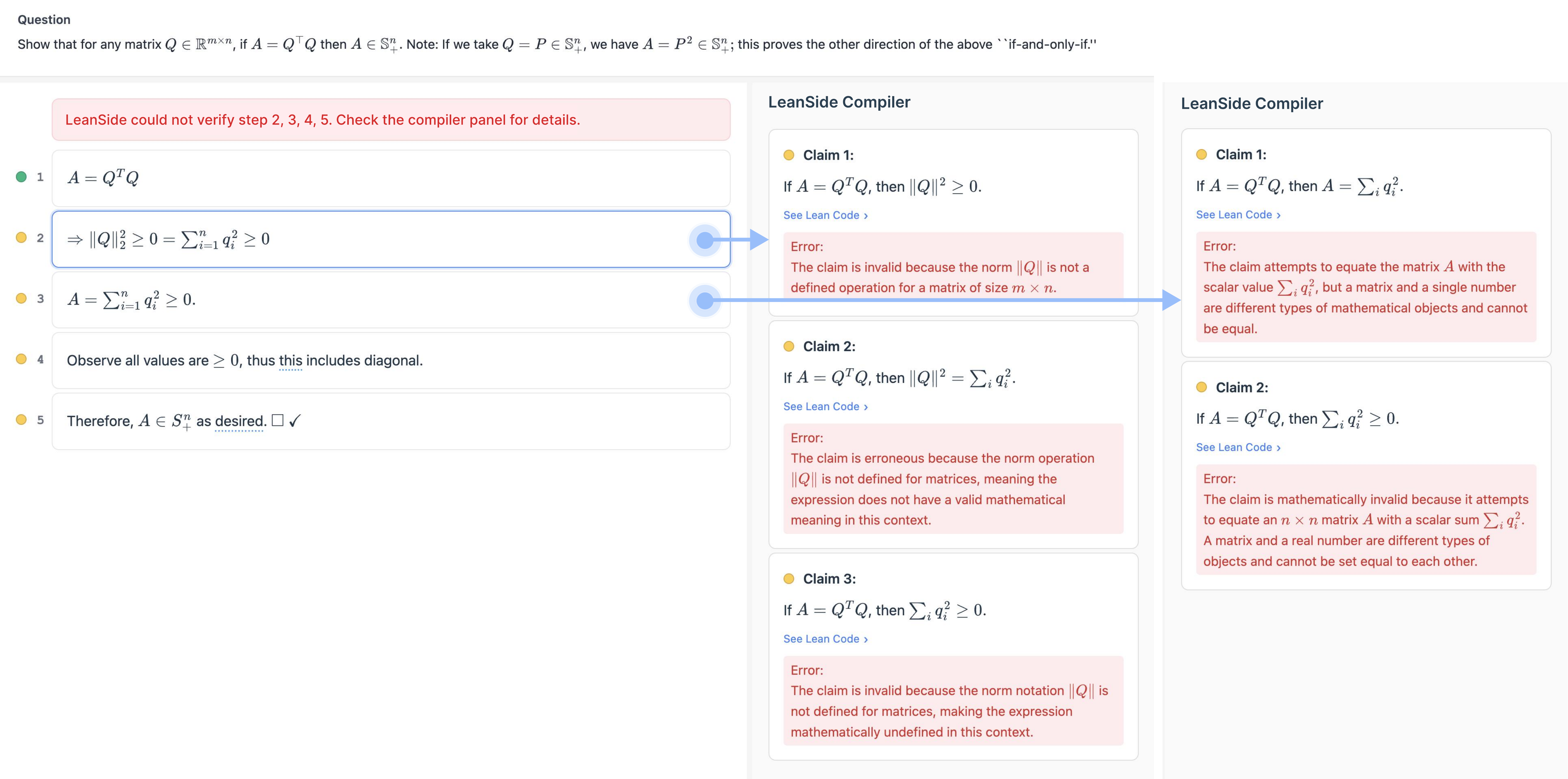}
    \vspace{-10pt}
    \caption{This figure shows an accurate but less understandable error messages of step 2 and 3.}
    \label{fig:eg2}
    \vspace{-10pt}
\end{figure}
\subsubsection{Progress type 3: Advancing with continuation hints and large-jump warnings (6 iterations)}

Students sometimes became stuck midway through an argument or did not know how to begin. They therefore entered given conditions or potentially relevant theorems as tentative starting points. When these preliminary steps were correct, the system could return informalized future proof steps as a continuation hint. As reported in Section~\ref{system_performance_understandbility}, these hints were generally understandable, and six students used them to make valid progress (Appendix \ref{appendix_more_examples}).

Unexpectedly, students also used large-jump warnings to advance incomplete proofs. This occurred when a student included the conclusion as the final step without providing a complete argument. Particularly for simpler problems, the prover could justify the conclusion within that step and return a large-jump warning exposing the missing intermediate reasoning. These warnings therefore functioned similarly to continuation hints. Appendix \ref{appendix_more_examples} display one such example.

Overall, students described next-step hints, including those presented as warnings, more understandable and helpful than continuation error messages. They described it as directly actionable: \textit{``It tells me how to do it''} (P4), \textit{``It's kind of funny that this seems a lot of help, seems like these are alternatives, not do all of them''} (P11), and \textit{``This is definitely very helpful''} (P18).

\subsubsection{Progress type 4: Discovering mistakes through unfaithful auto-corrections (11 iterations)}
\label{start_with_incorrect_proof}
As described in Section~\ref{system_performance_on_faithfulness}, the autoformalizer sometimes unfaithfully corrected local mistakes. Students discovered their error by noticing that the informalized verified proposition differed from their input step. These errors were typically small, such as omitting a transpose and thereby multiplying matrices with incompatible dimensions.

\subsubsection{Progress type 5: Resolving difficult feedback with external resources (11 iterations)}

Students could still progress when feedback was accurate but less understandable than human-tutor feedback (Section~\ref{system_performance_understandbility}), although doing so required more time and greater reliance on external resources such as browsers and textbooks.
For Q12 (Figure~\ref{fig:eg2}, P20 was confident that a matrix has a positive norm and therefore rejected the step~2 message with comments ``$\lVert Q\rVert$ is not a defined operation.'' P20 searched, \textit{``Can you take the norm of a non-square matrix?''} and treated an AI-generated overview beginning with \textit{``Yes, you can...''} as confirmation without reading further. However, the step~3 message---\textit{``a matrix and a single number cannot be equal''}---prompted P20 to reconsider. Returning to the search results, P20 read that the answer referred to specific norms, \textit{``...such as the Frobenius and induced operator norms,''} and recognized that the step~2 message was correct.

\begin{figure}[t]
    \centering
    \includegraphics[width=1\linewidth]{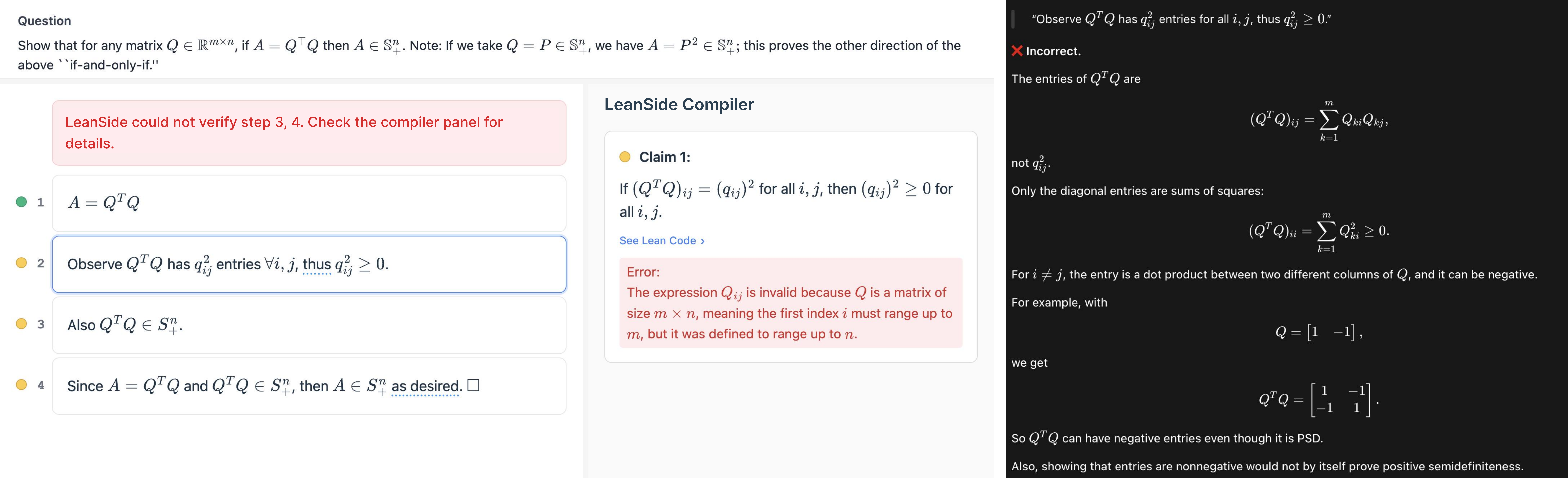}
    \caption{This figure shows a comparison between the LeanSide error message and an LLM output with the same problem and answer.}
    \vspace{-10pt}
    \label{fig:eg3}
\end{figure}
\subsection{When do students get stuck or divert from correct directions?} \label{stuck}
\subsubsection{Stuck type 1: Struggling with accurate but hard-to-understand feedback (16 cases)}
One common source of not understandable feedback, discussed in Section~\ref{system_performance_understandbility}, was a mismatch in granularity: Lean exposed low-level proof obligations for steps that students considered sufficiently explained. Students then struggled to connect the feedback to the higher-level flaw.
For example, Figure~\ref{fig:eg3} shows such an error message. P12 spent additional time searching lecture notes to understand the issue. By contrast, when given the same problem and solution, ChatGPT-4.0 identified the issue at a higher level.

In contrast, human tutors can draw on pedagogical strategies that our Lean-backed pipeline did not support, such as refuting an invalid claim with a counterexample~\citep{giannakoulias2010argumentation}, diagnosing an issue in the overall proof strategy rather than only the current step~\citep{autexier2012tutor}, or explaining an unfamiliar idea through an analogy to familiar knowledge~\citep{lulis2004analogy}.

\subsubsection{Stuck type 2: Lacking a high-level proof strategy (7 cases)} \label{stuck_lack_high_level}
Earlier, we showed that continuation hints and, less often, understandable continuation error messages could suggest plausible next steps. This support was less effective when students had no developed proof strategy and submitted reasoning far from a complete proof. In these cases, the proof assembler usually failed at an incidental proof obligation, producing a low-level error rather than the high-level strategy the student needed.
\begin{figure}[t]
    \centering
    \includegraphics[width=1\linewidth]{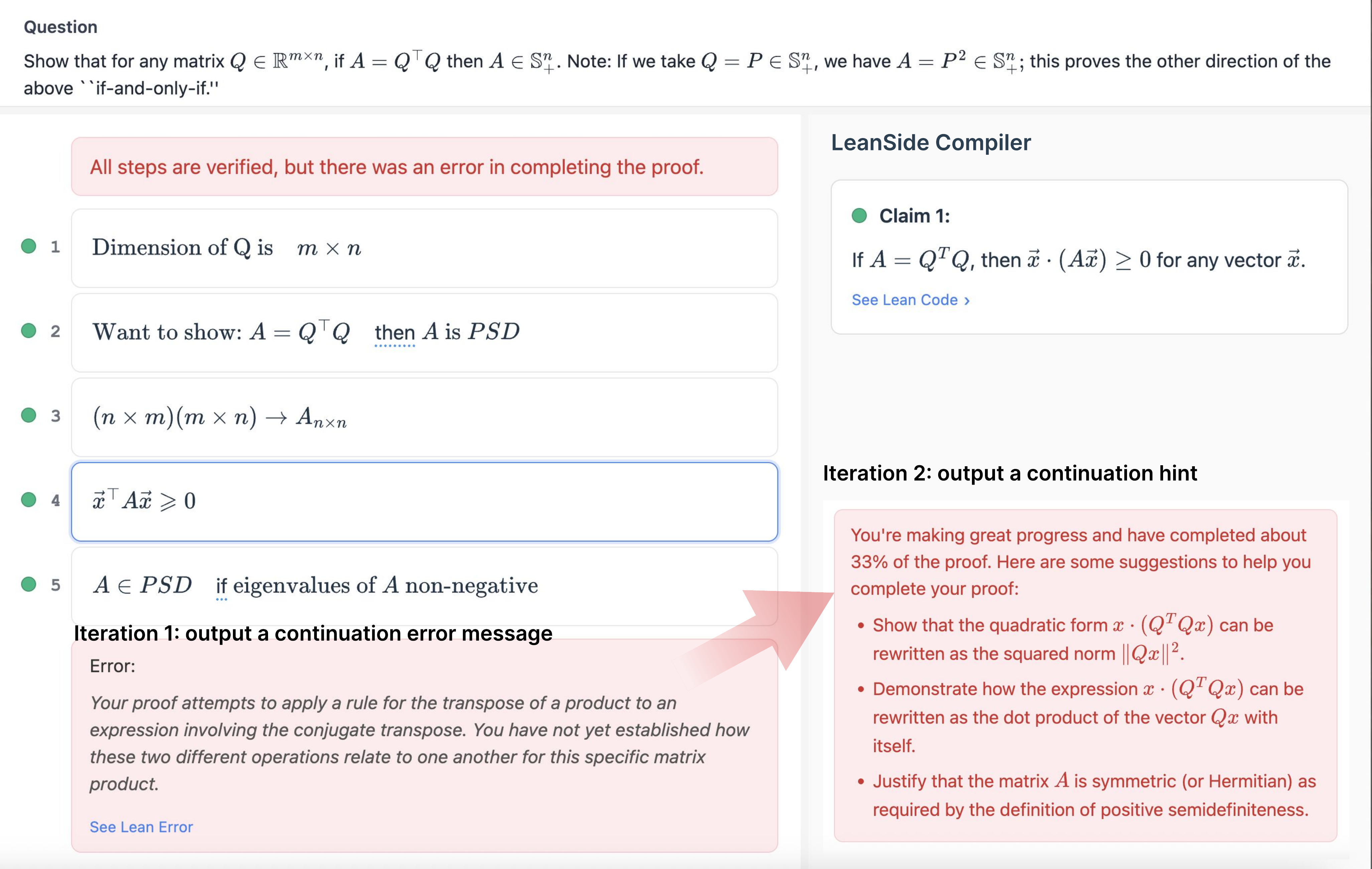}
    \caption{This figure shows a comparison between an unhelpful continuation error message and a useful continuation hint. In the first iteration, the system generated a confusing continuation error message; after the student re-verified the same proof, the system produced a useful continuation hint.}
    \vspace{-10pt}
    \label{fig:eg4}
\end{figure}
P16’s attempt on Q12 illustrates the contrast (Figure~\ref{fig:eg4}). After restating several given conditions, P16 became stuck. On the first run, the proof assembler failed and produced a continuation error mentioning a ``conjugate transpose.'' P16 rejected this message because the course considered only real numbers and reran the same partial proof. On the second run, the assembler completed the proof and returned a useful continuation hint, which enabled P16 to progress.

Additionally, the system could not distinguish strategically useful steps from all mathematically correct steps. Consequently, listing every potentially relevant fact was ineffective. This problem was compounded when a correct but irrelevant textbook fact fell outside the mini lemma library and was marked unverified, misleading students into thinking that the fact itself was incorrect.

\subsubsection{Diversion: Struggling to evaluate or act on local feedback (2 cases)}

The reason for students to drive into wrong direction is insufficient math knowledge to react on a step-wise error message no matter whether the message is accurate. For both cases where (1) the student's step is correct, however the system is inaccurate and failed to prove the step thus generate an error message, or (2) the student's step is indeed incorrect and thus the system generate an error message, it would requires student's knowledge to understand and act upon the error message. In two cases, Students followed the misleading feedback or misunderstand an accurate feedback.

\subsection {Participants gained confidence from the verified results}

A central motivation for formal verification is confidence in logical validity. We therefore assessed whether the system could carry this benefit into a natural-language interface by asking students to rate their confidence in each proof on a 1--5 scale and explain their reasons. We observed that when the system displayed a whole proof as verified for a proof that students were not confident in, their confidence did not immediately increase. Instead, confidence increased substantially as students checked whether each step matched their intended meaning, especially for steps they judged to be both faithfully translated and verified (P2, P4, P9, P11, P26). P4 commented, \textit{``I feel extremely confident that this is correct because it is verified,''}, P2 and P11 said they trusted the system when it marked a step as correct. P26 added, \textit{``I trust that my steps are rigorous for the verified claims more than LLM.''}

Students gradually gained trust through the process of checking that each informalized claim faithfully represented what they meant. Confidence increased especially when a step they had suspected might contain a gap was verified but accompanied by a large-jump warning or a continuation suggestion that clarified the missing details. For example, before verification, P9 rated their confidence as 2. Seeing that the proof was verified did not immediately increase this confidence: \textit{``Oh, it says it's correct, that's surprising.''} However, after checking that each step matched their intent and seeing that the uncertain step had a large-jump warning explaining the missing details, P9 became more confident in the proof. We also observed that P6, P9, and P13 became gradually more confident as they corrected mistakes or resolved ambiguous parts of their proofs.

\subsection{Could students obtain extra information from reading Lean code?}

To evaluate whether Lean code provided useful information beyond the interface, we invited students during the post-interview to read the generated Lean code. After a five-minute introduction to basic syntax, students read one proof lemma corresponding to one of their steps and could ask clarification to any part of the code. We then asked whether the code provided any helpful information not already shown in the interface.

Overall, 20 of 23 students said it was useful to know that code existed behind the feedback, but that reading the code did not give them additional information they needed. P4 commented that the Lean code felt unfamiliar and that \textit{``if it were just me, I would never open the panel and try to read the code.''} P11 said the language was daunting and that, even with instructions, they could mostly read only the comments. P14 similarly commented, \textit{``I trust the system more because I know there is code behind, but for now, I did not get much extra information from it.''}

The remaining three students found the comments inside the Lean lemmas useful. In all three cases, students had written steps that were verified but were not fully sure why the steps were correct. For P6, the system verified all steps, but P6 remained unsure about the evidence supporting one of the steps. P6 commented that the feedback \textit{``tells me it is correct, but I don't know how it is correct.''} After opening the Lean code panel, the comments helped P6 understand the evidence and gain confidence. P24 used the comments in the Lean Code to discover that a verified step was true only for a trivial reason: both sides reduced to \(0=0\), rather than for the reason the student intended. P7 used the comments to interpret a confusing large-jump warning: the step was circular because the prover had used the final conclusion to prove it. After reading the comments, the warning became understandable.
The comments in the Lean code represent a detailed logic trace below the user's granularity. These three cases show the potential usefulness of such traces and students' trust in them.

\begin{table*}[t]
    \centering
    \caption{Capabilities of different reasoning systems.}
    \label{tab:lean-backbone}
    \small
    \begin{tabularx}{0.8\textwidth}{@{}Xccc@{}}
        \toprule
        \textbf{Capability} &
        \textbf{LLM-based assistant} &
        \textbf{Formal debugger} &
        \textbf{Lean-backed system} \\
        \midrule
        Low-threshold interaction
            & $\checkmark$ & $\times$ & $\checkmark$ \\
        Preserve faithfulness and user agency
            & Not deterministic & $\checkmark$ & $\checkmark$ \\
        Validity check for informal reasoning
            & Not deterministic & $\times$ & $\checkmark$ \\
        Verified reasoning explanations
            & Not deterministic & $\times$ & $\checkmark$ \\
        Provide high-level conceptual guidance
            & $\checkmark$ & $\times$ & $\times$ \\
        \bottomrule
    \end{tabularx}
\end{table*}
\section{Discussion}
In the previous section, we summarized students' perceptions of the system: whether they could accurately assess system performance, when they were able to make progress, when they became stuck or diverted, and how their confidence changed. Here, we examine what explains these results. We distinguish findings that stem from current technical limitations from findings shaped by formally verified languages' properties and our design choices. We then discuss design implications for future formally verified co-reasoning (FVCR) systems, including the limitation of formal languages, the design strategies that may address them, and the open challenges that remain. Table~\ref{tab:lean-backbone} summarizes five capabilities that a co-reasoning system may need and compares our system with the support provided by an LLM-based assistant and a formal debugger.

\subsection{What results stem from technical limitations?}

LeanSide was developed under the cost constraints of a classroom deployment, so each verification was optimized for responsiveness and resource use. Recent commercial auto-formalizers~\cite{achim2025aristotle} report higher success rates in generating compiling proofs for correct Lean statements. Plugging stronger models into the pipeline would likely improve accuracy by reducing false negatives, and may also slightly reduce unfaithfulness. By contrast, we do not expect stronger provers to substantially change error-message understandability or most student behaviors around making progress or getting stuck, as these are more closely tied to the nature of formal verification and our system design choices.

\subsection{What could an FVCR system provide?}

We view an FVCR system like LeanSide could contributes four resources to a human-AI co-reasoning system: preserving the user's reasoning path, providing low-false-positive validity checks, producing hallucination-free reasoning feedback, and helping users iteratively make implicit reasoning explicit.

First, our findings show that LeanSide can help \textbf{preserve user agency through faithful representation}. Although we observed edge cases where the system unfaithfully auto-corrected step-level mistakes, LeanSide usually followed the user's reasoning at the user's granularity. This was not due to formal verification alone, but our system design choice for step-by-step auto-formalization. Prior work in human-AI collaboration has shown that LLM outputs can introduce useful directions, but can also unintentionally distract users from their original initiative~\citep{jakesch2023co,levy2021assessing}. Plausible but misaligned outputs can pull users away from their original line of reasoning, forcing them to monitor and repair the system or follow its suggested direction~\citep{subramonyam2024bridging,mozannar2024reading,vila2025miscommunicating}. An FVCR system can reduce this risk by designing the system to always follow the user's reasoning path. In our system, the unit of preservation is the pre-processed smallest progression unit: we do not require the system to preserve lower-level decompositions inside that unit. This unit is context-dependent and can be customized to the reasoning task. For example, in a tax-reasoning system, the unit might be one claimed inference: given a taxpayer's income, deductions, filing status, or transaction history, the system would preserve the user's intended application of a specific tax rule and then verify whether the stated tax consequence follows.

Second, our findings also shows an FVCR system can help \textbf{localize errors through low-false-positive feedback}. Hallucination remains a persistent concern in high-stakes reasoning domains. LLM feedback can be unreliable in these settings because it may sound plausible while being logically wrong~\cite{kim2025fostering, bo2025rely, spatharioti2025effects}. A formally verified backend can instead provide a stronger basis for accepting a step or conclusion when the translation is faithful. In our system, verified steps helped students narrow down the issue: once a step was verified, students gained confidence in its validity and could focus on unverified steps to correct faulty logic or fill the gaps. This process continued until all steps were verified, or until students used their mathematical knowledge to judge that a remaining unverified step reflected prover failure rather than their own error, as described in Section~\ref{system_performance_on_accuracy}.

Third, an FVCR system can provide \textbf{verified reasoning suggestions}. In our system, these traces appeared as continuation hints and as informal comments in the Lean code panel explaining how a step was proved. Both were natural-language explanations grounded in verified reasoning, and both were proven to be useful to students. Similar opportunities may arise in other reasoning tasks where users need more than a validity check mark: they also need natural-language explanations of why a step follows.

Finally, an FVCR system can help users make \textbf{implicit reasoning explicit}. In our study, students often entered high-level or ambiguous reasoning, while the system returned less ambiguous informalizations at the user's granularity. By comparing their input with these interpretations, students could notice ambiguity, clarify assumptions, and make their reasoning more rigorous. This suggests that formal verification can serve not only as a checker, but also as a mirror for helping users refine high-level strategies into precise reasoning.

\subsection{LeanSide Limitations, and Design Implications for Future FVCR Systems}
LeanSide was designed to isolate how formal language can support students' proof iteration. Some of its limitations therefore come from deliberate design choices, which could help explain why students became stuck or diverged.

First, Lean error messages can reflect \textbf{overly-granular reasoning}, which can confuse users. The limited understandability of step-level error messages illustrates this limitation. When students wrote a statement as a single assertion, they often treated it as not requiring further explanation in the course context. Lean errors, however, often reflected failures in lower-level proof obligations, making the feedback more granular than the student's step and harder to interpret. This mismatch is difficult to solve through the Lean backbone alone. Complementary methods, such as sampling counterexamples or using an LLM to generate higher-level pedagogical and tutor-like feedback, can help produce more useful explanations (formal verification could also be used to check whether the LLM generated feedback before showing it to users).

The second limitation is that \textbf{Lean may not follow the same strategic trace as human intuition}. In our study, students found the system useful for narrowing down local issues, but less able to guide them toward the right overall strategy. When a student followed an unproductive path, several mathematically correct steps could appear before the first incorrect step; the system flagged only the incorrect step, leaving the student to identify where the proof had diverged strategically. This reflects a mismatch between formal verification and human proof reasoning: Lean and search-based provers do not necessarily follow the intuitive order in which humans understand a proof, and may produce proofs that are counterintuitive or difficult to read~\citep{shi2025qed}. Strategic hint generation, such as tree-search-based methods studied in existing tutoring systems, could therefore complement the formal verification backbone.

Together, what FVCR systems can provide and where they are limited offers guidance for future co-reasoning tool designers on when to use an FVCR system. It also suggests that, when using a formal verification system as the backbone, task-specific auxiliary components can add useful ``meat'' around formal verification, varying by domain and interaction need.

\begin{figure}[t]
    \centering
    \includegraphics[width=1\linewidth]{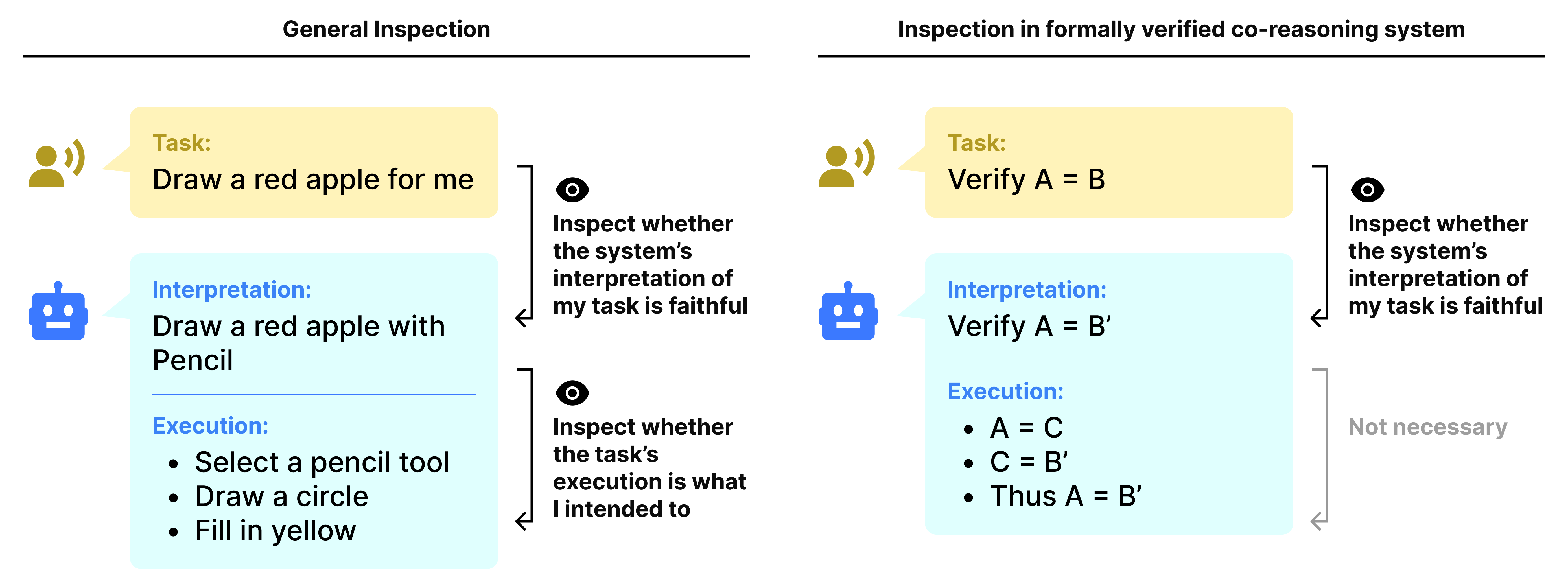}
    \caption{This figure compares two inspection scenarios: a general inspection scenario and inspection in an FVCR system. In the general scenario, users must inspect both interpretation and execution; in an FVCR system, the emphasis is primarily on inspecting interpretation.}
    \label{fig:inspection}
\end{figure}

\subsection{Open design challenges: faithfulness checking and auto-detecting contextual granularity}
\label{discussion_faithfulness}
One open design challenge is \textbf{faithfulness checking}. 
In figure \ref{fig:inspection}, we summarize users' inspection of a collaborative system into two categories (1) inspecting whether the system's interpretation of the presented task is faithful, and (2) inspecting whether the system's execution matches the user's intent, which could be correspond to gulf of execution and evaluation \cite{norman2013design}. We view inspection in FVCR systems as a special case of this broader inspection problem, one that emphasizes (1) while placing less burden on (2).

In LeanSide, the side-by-side comparison between user input and system interpretation in the informalization panel makes faithfulness inspection explicit. The open design challenge is how to make these checks lightweight enough to avoid tedious interaction, while still keeping them visible enough to hold users' attention.

We observed that students paid more attention to steps marked gray with error messages than to steps marked green, they did not always check whether the system had correctly interpreted what they meant in each sentence before everything is green. 

To address this issue, our current design requires users to click ``this is what I meant'' or ``not what I meant'' for every informalized cell before submitting the assignment. If a user clicks ``not what I meant,'' the step returns to gray and the student must clarify and re-verify the proof. However, this design has limitations. Students usually checked every step only after resolving all error messages and preparing to submit the assignment. As a result, participants sometimes ignored unfaithful interpretations during intermediate iterations. However, an unfaithful earlier step could prevent later causally related steps from being verified, increasing the likelihood of inaccurate judgments and error messages for the later step. This is one reason participants became stuck or moved in the wrong direction, as discussed in Section~\ref{stuck}.

\begin{figure}
    \centering
    \includegraphics[width=1\linewidth]{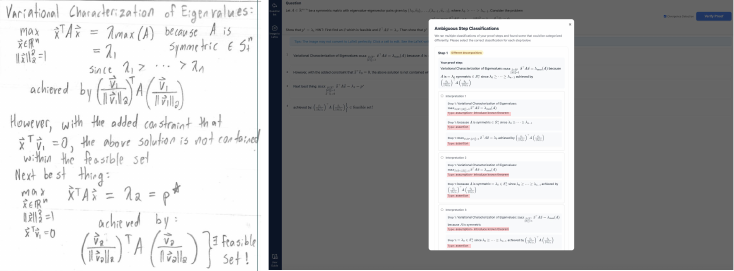}
    \caption{Follow up questions asked by the system when a student's step is ambiguous}
    \label{fig:follow_up}
    \vspace{-10pt}
\end{figure}

We considered two alternative designs. First, after each verification, the system could require students to confirm ``this is what I meant'' or ``not what I meant'' before showing indicators and error messages. However, because informalizations often differ from the student's original wording even when they are faithful, this interaction could become tedious. In addition, if a student clicked ``not what I meant,'' the system would need to regenerate the Lean interpretation, adding waiting time while leaving open the possibility that the new interpretation would still be unfaithful. Second, similar to existing LLM interfaces, the system could ask follow-up questions when the student's input is ambiguous. We implemented this feature in the interface shown in Figure \ref{fig:follow_up}, but because it adds complexity beyond the scope of the present study, we leave its evaluation to future work.

Another open design challenge is \textbf{auto-detecting contextual granularity}. Human users, such as students, often learn the acceptable granularity of reasoning from surrounding context, including lectures, peer discussion, and grading expectations, without anyone stating it explicitly. By contrast, LeanSide required substantial effort to construct a mini-lemma library that encoded this granularity, through rubric lemmas and side lemmas. Even with an automated pipeline, experts still had to inspect whether the specified granularity was appropriate. In mathematics education, this granularity is relatively visible because rubrics exist for each problem; in many other reasoning tasks, it may be far more implicit. A key open challenge for co-reasoning systems is therefore how they can learn the user's expected granularity for communicating reasoning.

\section{Limitations}

This study has several limitations. First, the participant population was relatively homogeneous: all participants were recruited from the same university, and many were majoring in Computer Science, Mathematics, or related fields. This composition was partly a consequence of the course in which we deployed the system. Although none of the participants had prior experience with Lean, they may have been more familiar with formal verification than the general population and therefore more inclined to trust the verification results. In addition, their disciplinary training may have helped them express informal reasoning more precisely, making their inputs less ambiguous than those of general users.

Second, the problems selected for this study represent only a limited subset of mathematical reasoning, focusing primarily on Linear Algebra and Optimization. This choice was motivated by two factors: (1) the deployment course focused on Linear Algebra, and (2) the current auto-formalization and prover modules perform more reliably in these domains. Consequently, the observed results may not generalize to other areas of mathematics, where users may exhibit different input behaviors and where the system may face additional challenges.

\newpage
\section{Acknowledgment}
The authors would like to acknowledge support from: NSF CAREER grant ECCS-2240031, the 2023 CITRIS Institute Seed Grant, the UC Berkeley Instructional Technology and Innovation MicroGrant Program, Google and the Renaissance Philanthropy AI for Math 2025 fund.
\bibliographystyle{ACM-Reference-Format}
\bibliography{references/icml, references/UIST,references/consolidated}

@article{clark1991grounding,
  title={Grounding in communication.},
  author={Clark, Herbert H and Brennan, Susan E},
  year={1991},
  publisher={American Psychological Association}
}

@article{clark1986referring,
  title={Referring as a collaborative process},
  author={Clark, Herbert H and Wilkes-Gibbs, Deanna},
  journal={Cognition},
  volume={22},
  number={1},
  pages={1--39},
  year={1986},
  publisher={Elsevier}
}

@article{myers2000past,
  title={Past, present, and future of user interface software tools},
  author={Myers, Brad and Hudson, Scott E and Pausch, Randy},
  journal={ACM Transactions on Computer-Human Interaction (TOCHI)},
  volume={7},
  number={1},
  pages={3--28},
  year={2000},
  publisher={ACM New York, NY, USA}
}

@article{resnick2005design,
  title={Design principles for tools to support creative thinking},
  author={Resnick, Mitchel and Myers, Brad and Nakakoji, Kumiyo and Shneiderman, Ben and Pausch, Randy and Selker, Ted and Eisenberg, Mike},
  year={2005},
  publisher={Carnegie Mellon University}
}

@incollection{tarski1936logical,
  author    = {Tarski, Alfred},
  title     = {On the Concept of Logical Consequence},
  booktitle = {Logic, Semantics, Metamathematics: Papers from 1923 to 1938},
  publisher = {Oxford University Press},
  year      = {1983},
  note      = {Original work published 1936}
}

@article{hendrycks2021measuring,
  title={Measuring mathematical problem solving with the math dataset},
  author={Hendrycks, Dan and Burns, Collin and Kadavath, Saurav and Arora, Akul and Basart, Steven and Tang, Eric and Song, Dawn and Steinhardt, Jacob},
  journal={arXiv preprint arXiv:2103.03874},
  year={2021}
}

@article{cobbe2021training,
  title={Training verifiers to solve math word problems, 2021},
  author={Cobbe, Karl and Kosaraju, Vineet and Bavarian, Mohammad and Chen, Mark and Jun, Heewoo and Kaiser, Lukasz and Plappert, Matthias and Tworek, Jerry and Hilton, Jacob and Nakano, Reiichiro and others},
  journal={URL https://arxiv. org/abs/2110.14168},
  volume={9},
  year={2021}
}

@article{wei2022chain,
  title={Chain-of-thought prompting elicits reasoning in large language models},
  author={Wei, Jason and Wang, Xuezhi and Schuurmans, Dale and Bosma, Maarten and Xia, Fei and Chi, Ed and Le, Quoc V and Zhou, Denny and others},
  journal={Advances in neural information processing systems},
  volume={35},
  pages={24824--24837},
  year={2022}
}

@article{lewkowycz2022solving,
  title={Solving quantitative reasoning problems with language models},
  author={Lewkowycz, Aitor and Andreassen, Anders and Dohan, David and Dyer, Ethan and Michalewski, Henryk and Ramasesh, Vinay and Slone, Ambrose and Anil, Cem and Schlag, Imanol and Gutman-Solo, Theo and others},
  journal={Advances in neural information processing systems},
  volume={35},
  pages={3843--3857},
  year={2022}
}

@article{li2022competition,
  title={Competition-level code generation with alphacode},
  author={Li, Yujia and Choi, David and Chung, Junyoung and Kushman, Nate and Schrittwieser, Julian and Leblond, R{\'e}mi and Eccles, Tom and Keeling, James and Gimeno, Felix and Dal Lago, Agustin and others},
  journal={Science},
  volume={378},
  number={6624},
  pages={1092--1097},
  year={2022},
  publisher={American Association for the Advancement of Science}
}

@article{guha2023legalbench,
  title={Legalbench: A collaboratively built benchmark for measuring legal reasoning in large language models},
  author={Guha, Neel and Nyarko, Julian and Ho, Daniel and R{\'e}, Christopher and Chilton, Adam and Chohlas-Wood, Alex and Peters, Austin and Waldon, Brandon and Rockmore, Daniel and Zambrano, Diego and others},
  journal={Advances in neural information processing systems},
  volume={36},
  pages={44123--44279},
  year={2023}
}

@article{singhal2023large,
  title={Large language models encode clinical knowledge},
  author={Singhal, Karan and Azizi, Shekoofeh and Tu, Tao and Mahdavi, S Sara and Wei, Jason and Chung, Hyung Won and Scales, Nathan and Tanwani, Ajay and Cole-Lewis, Heather and Pfohl, Stephen and others},
  journal={Nature},
  volume={620},
  number={7972},
  pages={172--180},
  year={2023},
  publisher={Nature Publishing Group UK London}
}

@article{kung2023performance,
  title={Performance of ChatGPT on USMLE: potential for AI-assisted medical education using large language models},
  author={Kung, Tiffany H and Cheatham, Morgan and Medenilla, Arielle and Sillos, Czarina and De Leon, Lorie and Elepa{\~n}o, Camille and Madriaga, Maria and Aggabao, Rimel and Diaz-Candido, Giezel and Maningo, James and others},
  journal={PLoS digital health},
  volume={2},
  number={2},
  pages={e0000198},
  year={2023},
  publisher={Public Library of Science}
}

@article{nori2023capabilities,
  title={Capabilities of gpt-4 on medical challenge problems},
  author={Nori, Harsha and King, Nicholas and McKinney, Scott Mayer and Carignan, Dean and Horvitz, Eric},
  journal={arXiv preprint arXiv:2303.13375},
  year={2023}
}

@article{xu2023large,
  title={Are large language models really good logical reasoners? a comprehensive evaluation and beyond},
  author={Xu, Fangzhi and Lin, Qika and Han, Jiawei and Zhao, Tianzhe and Liu, Jun and Cambria, Erik},
  journal={IEEE Transactions on Knowledge and Data Engineering},
  volume={37},
  number={4},
  pages={1620--1634},
  year={2025},
  publisher={IEEE}
}

@article{ji2023survey,
  title={Survey of hallucination in natural language generation},
  author={Ji, Ziwei and Lee, Nayeon and Frieske, Rita and Yu, Tiezheng and Su, Dan and Xu, Yan and Ishii, Etsuko and Bang, Ye Jin and Madotto, Andrea and Fung, Pascale},
  journal={ACM computing surveys},
  volume={55},
  number={12},
  pages={1--38},
  year={2023},
  publisher={ACM New York, NY}
}

@article{kasneci2023chatgpt,
  title={ChatGPT for good? On opportunities and challenges of large language models for education},
  author={Kasneci, Enkelejda and Se{\ss}ler, Kathrin and K{\"u}chemann, Stefan and Bannert, Maria and Dementieva, Daryna and Fischer, Frank and Gasser, Urs and Groh, Georg and G{\"u}nnemann, Stephan and H{\"u}llermeier, Eyke and others},
  journal={Learning and individual differences},
  volume={103},
  pages={102274},
  year={2023},
  publisher={Elsevier}
}

@inproceedings{wang2023examining,
  title={Examining the potential and pitfalls of ChatGPT in science and engineering problem-solving},
  author={Wang, Karen D and Burkholder, Eric and Wieman, Carl and Salehi, Shima and Haber, Nick},
  booktitle={Frontiers in Education},
  volume={8},
  pages={1330486},
  year={2024},
  organization={Frontiers Media SA}
}

@article{liu2023novice,
  title={Novice learner and expert tutor: Evaluating math reasoning abilities of large language models with misconceptions},
  author={Liu, Naiming and Sonkar, Shashank and Wang, Zichao and Woodhead, Simon and Baraniuk, Richard G},
  journal={arXiv preprint arXiv:2310.02439},
  year={2023}
}

@article{chen2026chat,
  title={Chat-based support alone may not be enough: comparing conversational and embedded LLM feedback for mathematical proof learning},
  author={Chen, Eason and Judicke, Sophia and Beigh, Kayla and Tang, Xinyi and Wang, Isabel and Yuan, Nina and Xiao, Zimo and Li, Chuangji and Li, Shizhuo and Luttmer, Reed and others},
  journal={arXiv preprint arXiv:2602.18807},
  year={2026}
}

@article{wardat2023chatgpt,
  title={ChatGPT: A revolutionary tool for teaching and learning mathematics},
  author={Wardat, Yousef},
  journal={Available at SSRN 5653030},
  year={2023}
}

@inproceedings{guo2026bridging,
  title={Bridging the gap to autoformalize free-form human reasoning},
  author={Guo, Chenjun and Patel, Manooshree and Mehta, Arnav and Bhattacharyya, Rayna and Lu, Thomas and Voss, Niels and Hartmann, Bjoern and Zamfirescu-Pereira, JD and Chasins, Sarah and Ranade, Gireeja},
  organization={Plateau Workshop}
}

@article{witharana2022survey,
  title={A survey on assertion-based hardware verification},
  author={Witharana, Hasini and Lyu, Yangdi and Charles, Subodha and Mishra, Prabhat},
  journal={ACM Computing Surveys (CSUR)},
  volume={54},
  number={11s},
  pages={1--33},
  year={2022},
  publisher={ACM New York, NY}
}

@inproceedings{paulson2023large,
  title={Large-scale formal proof for the working mathematician—lessons learnt from the ALEXANDRIA Project},
  author={Paulson, Lawrence C},
  booktitle={International Conference on Intelligent Computer Mathematics},
  pages={3--15},
  year={2023},
  organization={Springer}
}

@article{shulman2024strange,
  title={Strange new universes: Proof assistants and synthetic foundations},
  author={Shulman, Michael},
  journal={Bulletin of the American Mathematical Society},
  volume={61},
  number={2},
  pages={257--270},
  year={2024}
}

@article{weng2025autoformalization,
  title={Autoformalization in the era of large language models: A survey},
  author={Weng, Ke and Du, Lun and Li, Sirui and Lu, Wangyue and Sun, Haozhe and Liu, Hengyu and Zhang, Tiancheng},
  journal={arXiv preprint arXiv:2505.23486},
  year={2025}
}

@inproceedings{poiroux2025reliable,
  title={Reliable evaluation and benchmarks for statement autoformalization},
  author={Poiroux, Auguste and Weiss, Gail and Kun{\v{c}}ak, Viktor and Bosselut, Antoine},
  booktitle={Proceedings of the 2025 Conference on Empirical Methods in Natural Language Processing},
  pages={17958--17980},
  year={2025}
}

@article{cornish2026faithformbench,
  title={FaithformBench: Benchmarking Faithfulness of Mathematical Chain-of-Thought Autoformalisation},
  author={Cornish, Rob and Ghinassi, Iacopo and Yeh, Po-Hung and Liu, Shuqi and Xu, Qiyuan and Yin, Haoxuan and Wagner, Dominik and Li, Wenda and Teh, Yee Whye and Ong, Luke},
  journal={arXiv preprint arXiv:2608.10916},
  year={2026}
}

@article{zheng2021minif2f,
  title={Minif2f: a cross-system benchmark for formal olympiad-level mathematics},
  author={Zheng, Kunhao and Han, Jesse Michael and Polu, Stanislas},
  journal={arXiv preprint arXiv:2109.00110},
  year={2021}
}

@inproceedings{wolska2004analysis,
  title={Analysis of mixed natural and symbolic input in mathematical dialogs},
  author={Wolska, Magdalena and Kruijff-Korbayov{\'a}, Ivana},
  booktitle={Proceedings of the 42nd Annual Meeting of the Association for Computational Linguistics (ACL-04)},
  pages={25--32},
  year={2004}
}

@article{vanlehn2006behavior,
  title={The behavior of tutoring systems},
  author={VanLehn, Kurt},
  journal={International journal of artificial intelligence in education},
  volume={16},
  number={3},
  pages={227--265},
  year={2006},
  publisher={SAGE Publications Sage UK: London, England}
}

@inproceedings{benzmueller2007dialog,
  title={Natural language dialog with a tutor system for mathematical proofs},
  author={Benzm{\"u}ller, Christoph and Horacek, Helmut and Kruijff-Korbayova, Ivana and Pinkal, Manfred and Siekmann, J{\"o}rg and Wolska, Magdalena},
  booktitle={Cognitive Systems: Joint Chinese-German Workshop, Shanghai, China, March 7-11, 2005, Revised Selected Papers},
  pages={1--14},
  year={2007},
  organization={Springer}
}

@book{hanna2019proof,
  title={Proof technology in mathematics research and teaching},
  author={Hanna, Gila and Reid, David A and De Villiers, Michael},
  year={2019},
  publisher={Springer}
}

@article{hendriks2010teaching,
  title={Teaching logic using a state-of-the-art proof assistant.},
  author={Hendriks, Maxim and Kaliszyk, Cezary and Van Raamsdonk, Femke and Wiedijk, Freek},
  journal={Acta Didactica Napocensia},
  volume={3},
  number={2},
  pages={35--48},
  year={2010},
  publisher={ERIC}
}

@inproceedings{henz2011teaching,
  title={Teaching experience: Logic and formal methods with Coq},
  author={Henz, Martin and Hobor, Aquinas},
  booktitle={International Conference on Certified Programs and Proofs},
  pages={199--215},
  year={2011},
  organization={Springer}
}

@article{jacobsen2023exams,
  title={On exams with the Isabelle proof assistant},
  author={Jacobsen, Frederik Krogsdal and Villadsen, J{\o}rgen},
  journal={arXiv preprint arXiv:2303.05866},
  year={2023}
}

@article{thoma2022learning,
  title={Learning about proof with the theorem prover lean: the abundant numbers task},
  author={Thoma, Athina and Iannone, Paola},
  journal={International Journal of Research in Undergraduate Mathematics Education},
  volume={8},
  number={1},
  pages={64--93},
  year={2022},
  publisher={Springer}
}

@inproceedings{aldrich2008sasylf,
  title={SASyLF: An educational proof assistant for language theory},
  author={Aldrich, Jonathan and Simmons, Robert J and Shin, Key},
  booktitle={Proceedings of the 2008 international workshop on Functional and declarative programming in education},
  pages={31--40},
  year={2008}
}

@inproceedings{nawrocki2023extensible,
  title={An extensible user interface for Lean 4},
  author={Nawrocki, Wojciech and Ayers, Edward W and Ebner, Gabriel},
  booktitle={14th International Conference on Interactive Theorem Proving (ITP 2023)},
  year={2023},
  organization={Schloss Dagstuhl-Leibniz-Zentrum f{\"u}r Informatik}
}

@inproceedings{shi2025designing,
  title={Designing Proof Deautomation for Rocq},
  author={Shi, Jessica and Torczon, Cassia and Goldstein, Harrison and Head, Andrew and Pierce, Benjamin C},
  booktitle={Proceedings of the 15th PLATEAU Workshop on Programming Languages and Human-Computer Interaction (PLATEAU’25). ACM},
  year={2025}
}

@inproceedings{massot2024teaching,
  title={Teaching mathematics using lean and controlled natural language},
  author={Massot, Patrick},
  booktitle={15th International Conference on Interactive Theorem Proving (ITP 2024)},
  pages={27--1},
  year={2024},
  organization={Schloss Dagstuhl--Leibniz-Zentrum f{\"u}r Informatik}
}

@article{nelson2024forge,
  title={Forge: A tool and language for teaching formal methods},
  author={Nelson, Tim and Greenman, Ben and Prasad, Siddhartha and Dyer, Tristan and Bove, Ethan and Chen, Qianfan and Cutting, Charles and Del Vecchio, Thomas and LeVine, Sidney and Rudner, Julianne and others},
  journal={Proceedings of the ACM on Programming Languages},
  volume={8},
  number={OOPSLA1},
  pages={613--641},
  year={2024},
  publisher={ACM New York, NY, USA}
}

@misc{lea2026,
  title        = {{Lea}: An Agent Backbone for Mathematician-Led Formalization},
  author       = {{VIDA-NYU}},
  year         = {2026},
  howpublished = {\url{https://vida-nyu.github.io/Lea/}},
  note         = {Accessed: 2026-08-29}
}

@article{song2024lean,
  title={Lean copilot: Large language models as copilots for theorem proving in lean},
  author={Song, Peiyang and Yang, Kaiyu and Anandkumar, Anima},
  journal={arXiv preprint arXiv:2404.12534},
  year={2024}
}

@inproceedings{zamfirescu2023johnny,
  author    = {Zamfirescu-Pereira, J. Diego and Wong, Richmond Y. and
               Hartmann, Bjoern and Yang, Qian},
  title     = {Why Johnny Can't Prompt: How Non-{AI} Experts Try
               (and Fail) to Design {LLM} Prompts},
  booktitle = {Proceedings of the 2023 CHI Conference on Human Factors
               in Computing Systems},
  year      = {2023},
  articleno = {437},
  numpages  = {21},
  doi       = {10.1145/3544548.3581388}
}

@inproceedings{nguyen2024beginning,
  author    = {Nguyen, Sydney and Babe, Hannah McLean and Zi, Yangtian and
               Guha, Arjun and Anderson, Carolyn Jane and Feldman, Molly Q.},
  title     = {How Beginning Programmers and Code {LLM}s (Mis)read Each Other},
  booktitle = {Proceedings of the 2024 CHI Conference on Human Factors
               in Computing Systems},
  year      = {2024},
  pages     = {1--26},
  doi       = {10.1145/3613904.3642706}
}

@inproceedings{jiang2022discovering,
  author    = {Jiang, Ellen and Toh, Edwin and Molina, Alejandra and
               Olson, Kristen and Kayacik, Claire and Donsbach, Aaron and
               Cai, Carrie J. and Terry, Michael},
  title     = {Discovering the Syntax and Strategies of Natural Language
               Programming with Generative Language Models},
  booktitle = {Proceedings of the 2022 CHI Conference on Human Factors
               in Computing Systems},
  year      = {2022},
  pages     = {1--19},
  doi       = {10.1145/3491102.3501870}
}

@inproceedings{zi2025written,
  author    = {Zi, Yangtian and Li, Luisa and Guha, Arjun and
               Anderson, Carolyn Jane and Feldman, Molly Q.},
  title     = {{``I Would Have Written My Code Differently''}:
               Beginners Struggle to Understand {LLM}-Generated Code},
  booktitle = {Proceedings of the 33rd ACM International Conference
               on the Foundations of Software Engineering},
  pages     = {1479--1488},
  year      = {2025},
  doi       = {10.1145/3696630.3731663}
}

@inproceedings{subramonyam2024bridging,
  title={Bridging the gulf of envisioning: Cognitive challenges in prompt based interactions with LLMs},
  author={Subramonyam, Hari and Pea, Roy and Pondoc, Christopher and Agrawala, Maneesh and Seifert, Colleen},
  booktitle={Proceedings of the 2024 CHI Conference on Human Factors in Computing Systems},
  pages={1--19},
  year={2024}
}

@inproceedings{vila2025miscommunicating,
  author    = {Cros Vila, Laura and Sturm, Bob},
  title     = {(Mis)Communicating with our {AI} Systems},
  booktitle = {Proceedings of the 2025 CHI Conference on Human Factors
               in Computing Systems},
  year      = {2025},
  articleno = {416},
  numpages  = {9},
  doi       = {10.1145/3706598.3713771}
}

@article{barke2023grounded,
  title={Grounded copilot: How programmers interact with code-generating models},
  author={Barke, Shraddha and James, Michael B and Polikarpova, Nadia},
  journal={Proceedings of the ACM on Programming Languages},
  volume={7},
  number={OOPSLA1},
  pages={85--111},
  year={2023},
  publisher={ACM New York, NY, USA}
}

@inproceedings{levy2021assessing,
  title={Assessing the impact of automated suggestions on decision making: Domain experts mediate model errors but take less initiative},
  author={Levy, Ariel and Agrawal, Monica and Satyanarayan, Arvind and Sontag, David},
  booktitle={Proceedings of the 2021 CHI conference on human factors in computing systems},
  pages={1--13},
  year={2021}
}

@inproceedings{reicherts2025ai,
  title={AI, help me think—but for myself: Assisting people in complex decision-making by providing different kinds of cognitive support},
  author={Reicherts, Leon and Zhang, Zelun Tony and Von Oswald, Elisabeth and Liu, Yuanting and Rogers, Yvonne and Hassib, Mariam},
  booktitle={Proceedings of the 2025 CHI Conference on Human Factors in Computing Systems},
  pages={1--19},
  year={2025}
}

@article{draxler2024ghostwriter,
  title={The AI ghostwriter effect: When users do not perceive ownership of AI-generated text but self-declare as authors},
  author={Draxler, Fiona and Werner, Anna and Lehmann, Florian and Hoppe, Matthias and Schmidt, Albrecht and Buschek, Daniel and Welsch, Robin},
  journal={ACM Transactions on Computer-Human Interaction},
  volume={31},
  number={2},
  pages={1--40},
  year={2024},
  publisher={ACM New York, NY}
}

@inproceedings{jakesch2023co,
  title={Co-writing with opinionated language models affects users’ views},
  author={Jakesch, Maurice and Bhat, Advait and Buschek, Daniel and Zalmanson, Lior and Naaman, Mor},
  booktitle={Proceedings of the 2023 CHI conference on human factors in computing systems},
  pages={1--15},
  year={2023}
}

@inproceedings{liu2023wants,
  author    = {Liu, Michael Xieyang and Sarkar, Advait and Negreanu, Carina and
               Zorn, Benjamin and Williams, Jack and Toronto, Neil and
               Gordon, Andrew D.},
  title     = {{``What It Wants Me To Say''}: Bridging the Abstraction Gap
               Between End-User Programmers and Code-Generating Large
               Language Models},
  booktitle = {Proceedings of the 2023 CHI Conference on Human Factors
               in Computing Systems},
  year      = {2023},
  pages     = {1--31},
  doi       = {10.1145/3544548.3580817}
}

@inproceedings{li2020multi,
  author    = {Li, Toby Jia-Jun and Chen, Jingya and Xia, Haijun and
               Mitchell, Tom M. and Myers, Brad A.},
  title     = {Multi-Modal Repairs of Conversational Breakdowns
               in Task-Oriented Dialogs},
  booktitle = {Proceedings of the 33rd Annual ACM Symposium on
               User Interface Software and Technology},
  pages     = {1094--1107},
  year      = {2020},
  doi       = {10.1145/3379337.3415820}
}

@inproceedings{kim2025fostering,
  title={Fostering appropriate reliance on large language models: The role of explanations, sources, and inconsistencies},
  author={Kim, Sunnie SY and Vaughan, Jennifer Wortman and Liao, Q Vera and Lombrozo, Tania and Russakovsky, Olga},
  booktitle={Proceedings of the 2025 CHI Conference on Human Factors in Computing Systems},
  pages={1--19},
  year={2025}
}

@inproceedings{bo2025rely,
  title={To rely or not to rely? evaluating interventions for appropriate reliance on large language models},
  author={Bo, Jessica Y and Wan, Sophia and Anderson, Ashton},
  booktitle={Proceedings of the 2025 CHI Conference on Human Factors in Computing Systems},
  pages={1--23},
  year={2025}
}

@inproceedings{o2025scientists,
  title={How scientists use large language models to program},
  author={O'Brien, Gabrielle},
  booktitle={Proceedings of the 2025 CHI Conference on Human Factors in Computing Systems},
  pages={1--16},
  year={2025}
}

@inproceedings{cheng2024relic,
  title={Relic: Investigating large language model responses using self-consistency},
  author={Cheng, Furui and Zouhar, Vil{\'e}m and Arora, Simran and Sachan, Mrinmaya and Strobelt, Hendrik and El-Assady, Mennatallah},
  booktitle={Proceedings of the 2024 CHI conference on human factors in computing systems},
  pages={1--18},
  year={2024}
}

@inproceedings{leiser2024hill,
  title={Hill: A hallucination identifier for large language models},
  author={Leiser, Florian and Eckhardt, Sven and Leuthe, Valentin and Knaeble, Merlin and M{\"a}dche, Alexander and Schwabe, Gerhard and Sunyaev, Ali},
  booktitle={Proceedings of the 2024 CHI conference on human factors in computing systems},
  pages={1--13},
  year={2024}
}

@inproceedings{ferdowsi2024validating,
  title={Validating AI-generated code with live programming},
  author={Ferdowsi, Kasra and Huang, Ruanqianqian and James, Michael B and Polikarpova, Nadia and Lerner, Sorin},
  booktitle={Proceedings of the 2024 CHI conference on human factors in computing systems},
  pages={1--8},
  year={2024}
}

@inproceedings{spatharioti2025effects,
  title={Effects of LLM-based search on decision making: Speed, accuracy, and overreliance},
  author={Spatharioti, Sofia Eleni and Rothschild, David and Goldstein, Daniel G and Hofman, Jake M},
  booktitle={Proceedings of the 2025 CHI Conference on Human Factors in Computing Systems},
  pages={1--15},
  year={2025}
}

@article{shi2025qed,
  title={QED in context: An observation study of proof assistant users},
  author={Shi, Jessica and Torczon, Cassia and Goldstein, Harrison and Pierce, Benjamin C and Head, Andrew},
  journal={Proceedings of the ACM on Programming Languages},
  volume={9},
  number={OOPSLA1},
  pages={337--363},
  year={2025},
  publisher={ACM New York, NY, USA}
}

@misc{achim2025aristotle,
title         = {Aristotle: {IMO}-Level Automated Theorem Proving},
author        = {Achim, Tudor and Best, Alex and Bietti, Alberto and
Der, Kevin and F{'e}d{'e}rico, Math{"i}s and
Gukov, Sergei and Halpern-Leistner, Daniel and
Henningsgard, Kirsten and Kudryashov, Yury and
Meiburg, Alexander and Michelsen, Martin and
Patterson, Riley and Rodriguez, Eric and
Scharff, Laura and Shanker, Vikram and
Sicca, Vladmir and Sowrirajan, Hari and
Swope, Aidan and Tamas, Matyas and Tenev, Vlad and
Thomm, Jonathan and Williams, Harold and Wu, Lawrence},
year          = {2025},
eprint        = {2510.01346},
archivePrefix = {arXiv},
primaryClass  = {cs.AI},
doi           = {10.48550/arXiv.2510.01346},
url           = {https://arxiv.org/abs/2510.01346}
}

@misc{buzzard2023naturalnumbergame,
author       = {Buzzard, Kevin and Eugster, Jon},
title        = {The Natural Number Game},
year         = {2023},
howpublished = {Interactive Lean 4 game},
url          = {https://adam.math.hhu.de/\#/g/leanprover-community/NNG4},
note         = {Accessed: 2026-09-11}
}

@book{norman2013design,
  title={The design of everyday things: Revised and expanded edition},
  author={Norman, Don},
  year={2013},
  publisher={Basic books}
}

@article{giannakoulias2010argumentation,
  author  = {Giannakoulias, Eusthathios and Mastorides, Eleutherios
             and Potari, Despina and Zachariades, Theodossios},
  title   = {Studying Teachers' Mathematical Argumentation in the
             Context of Refuting Students' Invalid Claims},
  journal = {The Journal of Mathematical Behavior},
  volume  = {29},
  number  = {3},
  pages   = {160--168},
  year    = {2010},
  doi     = {10.1016/j.jmathb.2010.07.001}
}

@article{autexier2012tutor,
  author  = {Autexier, Serge and Dietrich, Dominik and Schiller, Marvin},
  title   = {Towards an Intelligent Tutor for Mathematical Proofs},
  journal = {Electronic Proceedings in Theoretical Computer Science},
  volume  = {79},
  pages   = {1--28},
  year    = {2012},
  doi     = {10.4204/EPTCS.79.1}
}

@inproceedings{lulis2004analogy,
  author    = {Lulis, Evelyn and Evens, Martha and Michael, Joel},
  title     = {How Human Tutors Employ Analogy to Facilitate Understanding},
  booktitle = {Proceedings of the Twenty-Sixth Annual Conference of the
               Cognitive Science Society},
  editor    = {Forbus, Kenneth and Gentner, Dedre and Regier, Terry},
  pages     = {861--866},
  year      = {2004},
  publisher = {Lawrence Erlbaum Associates}
}

@misc{lau2026aristotlecase,
  author        = {Lau, Gabriel Rongyang},
  title         = {Using the {Aristotle API} for {AI}-Assisted
                   Theorem Proving in {Lean 4}: A Formalisation
                   Case Study of the Grasshopper Problem},
  year          = {2026},
  eprint        = {2605.20120},
  archivePrefix = {arXiv},
  url           = {https://arxiv.org/abs/2605.20120}
}

@inproceedings{mozannar2024reading,
  author    = {Mozannar, Hussein and Bansal, Gagan and
               Fourney, Adam and Horvitz, Eric},
  title     = {Reading Between the Lines: Modeling User Behavior
               and Costs in {AI}-Assisted Programming},
  booktitle = {Proceedings of the 2024 CHI Conference on Human
               Factors in Computing Systems},
  year      = {2024},
  pages     = {1--16},
  publisher = {ACM},
  doi       = {10.1145/3613904.3641936}
}

@misc{openai2026navierstokes,
  author = {{OpenAI}},
  title = {On the {Navier--Stokes} Millennium Prize Problem},
  year = {2026},
  howpublished = {\url{https://openai.com/index/navier-stokes-solution/}},
  note = {Accessed September 11, 2026}
}

@misc{deepmind2024alphaproof,
  author = {{Google DeepMind}},
  title = {{AI} Achieves Silver-Medal Standard Solving International Mathematical Olympiad Problems},
  year = {2024},
  howpublished = {\url{https://deepmind.google/blog/ai-solves-imo-problems-at-silver-medal-level/}},
  note = {Accessed September 11, 2026}
}

@inproceedings{paulson1994isabelle,
  author    = {Lawrence C. Paulson},
  title     = {Isabelle: A Generic Theorem Prover},
  booktitle = {Proceedings of the 12th International Conference on Automated Deduction (CADE-12)},
  editor    = {Alan Bundy},
  series    = {Lecture Notes in Computer Science},
  volume    = {814},
  pages     = {37--41},
  publisher = {Springer},
  year      = {1994}
}

@inproceedings{deMouraUllrich2021lean4,
  author    = {Leonardo de Moura and Sebastian Ullrich},
  title     = {The {Lean} 4 Theorem Prover and Programming Language},
  booktitle = {International Conference on Automated Deduction},
  pages     = {625--635},
  publisher = {Springer},
  year      = {2021}
}

@techreport{barras1997coq,
  author      = {Bruno Barras and Samuel Boutin and Cristina Cornes and Judica{\"e}l Courant and Jean-Christophe Filli{\^a}tre and Eduardo Gim{\'e}nez and H{\'e}ctor Herbelin and G{\'e}rard Mohring and Amokrane Sa{\"i}bi and Benjamin Werner},
  title       = {The {Coq} Proof Assistant Reference Manual: Version 6.1},
  institution = {INRIA},
  number      = {RT-0203},
  year        = {1997}
}

@misc{liu2025safeenhancingmathematicalreasoning,
      title={Safe: Enhancing Mathematical Reasoning in Large Language Models via Retrospective Step-aware Formal Verification}, 
      author={Chengwu Liu and Ye Yuan and Yichun Yin and Yan Xu and Xin Xu and Zaoyu Chen and Yasheng Wang and Lifeng Shang and Qun Liu and Ming Zhang},
      year={2025},
      eprint={2506.04592},
      archivePrefix={arXiv},
      primaryClass={cs.CL},
      url={https://arxiv.org/abs/2506.04592}, 
}

@inproceedings{patel2026leantutorverifiedaimathematical,
  title     = {LeanTutor: Towards a Verified {AI} Mathematical Proof Tutor},
  author    = {Patel, Manooshree and Bhattacharyya, Rayna and Lu, Thomas and
               Mehta, Arnav and Voss, Niels and Norouzi, Narges and Ranade, Gireeja},
  booktitle = {Proceedings of the Fortieth AAAI Conference on Artificial Intelligence},
  series    = {AAAI~'26},
  pages     = {40670-40678},
  year      = {2026},
  month     = {January},
  publisher = {AAAI Press},
  url       = {https://ojs.aaai.org/index.php/AAAI/article/view/41514},
}

@inproceedings{
liu2025rethinking,
title={Rethinking and Improving Autoformalization: Towards a Faithful Metric and a Dependency Retrieval-based Approach},
author={Qi Liu and Xinhao Zheng and Xudong Lu and Qinxiang Cao and Junchi Yan},
booktitle={The Thirteenth International Conference on Learning Representations},
year={2025},
url={https://openreview.net/forum?id=hUb2At2DsQ}
}

@misc{jurayj2026languagemodelslogicprograms,
      title={Language Models and Logic Programs for Trustworthy Tax Reasoning}, 
      author={William Jurayj and Nils Holzenberger and Benjamin Van Durme},
      year={2026},
      eprint={2508.21051},
      archivePrefix={arXiv},
      primaryClass={cs.CL},
      url={https://arxiv.org/abs/2508.21051}, 
}

@article{shi2025qedobsstudy,
  author       = {Jessica Shi and Cassia Torczon and Harrison Goldstein and Benjamin C. Pierce and Andrew Head},
  title        = {QED in Context: An Observation Study of Proof Assistant Users},
  journal      = {Proceedings of the ACM on Programming Languages},
  volume       = {9},
  number       = {OOPSLA1},
  articleno    = {92},
  month        = apr,
  year         = {2025},
  pages        = {1--27},
  doi          = {10.1145/3720426},
  publisher    = {ACM}
}

@article{liu2025atlas,
  title={Atlas: Autoformalizing theorems through lifting, augmentation, and synthesis of data},
  author={Liu, Xiaoyang and Bao, Kangjie and Zhang, Jiashuo and Liu, Yunqi and Chen, Yu and Liu, Yuntian and Jiao, Yang and Luo, Tao},
  journal={arXiv preprint arXiv:2502.05567},
  year={2025}
}

@article{lu2024process,
  title={Process-driven autoformalization in lean 4},
  author={Lu, Jianqiao and Wan, Yingjia and Liu, Zhengying and Huang, Yinya and Xiong, Jing and Liu, Chengwu and Shen, Jianhao and Jin, Hui and Zhang, Jipeng and Wang, Haiming and others},
  journal={arXiv preprint arXiv:2406.01940},
  year={2024}
}

@inproceedings{wu2026stepfun,
  title={StepFun-Formalizer: Unlocking the Autoformalization Potential of LLMs Through Knowledge-Reasoning Fusion},
  author={Wu, Yutong and Huang, Di and Wan, Ruosi and Peng, Yue and Shang, Shijie and Cao, Chenrui and Qi, Lei and Zhang, Rui and Zhang, Xishan and Du, Zidong and others},
  booktitle={Proceedings of the AAAI Conference on Artificial Intelligence},
  volume={40},
  number={40},
  pages={33980--33988},
  year={2026}
}

@article{lin2025goedel,
  title={Goedel-prover-v2: Scaling formal theorem proving with scaffolded data synthesis and self-correction},
  author={Lin, Yong and Tang, Shange and Lyu, Bohan and Yang, Ziran and Chung, Jui-Hui and Zhao, Haoyu and Jiang, Lai and Geng, Yihan and Ge, Jiawei and Sun, Jingruo and others},
  journal={arXiv preprint arXiv:2508.03613},
  year={2025}
}

@article{cabral2025proofflow,
  title={ProofFlow: A Dependency Graph Approach to Faithful Proof Autoformalization},
  author={Cabral, Rafael and Do, Tuan Manh and Yu, Xuejun and Tai, Wai Ming and Feng, Zijin and Shen, Xin},
  journal={arXiv preprint arXiv:2510.15981},
  year={2025}
}

@article{murphy2024autoformalizing,
  title={Autoformalizing euclidean geometry},
  author={Murphy, Logan and Yang, Kaiyu and Sun, Jialiang and Li, Zhaoyu and Anandkumar, Anima and Si, Xujie},
  journal={arXiv preprint arXiv:2405.17216},
  year={2024}
}

@inproceedings{guo2025unspoken,
  title={Unspoken Logic: Understanding and bridging the gap between free-form and LLM-interpretable natural language mathematical proofs},
  author={Guo, Chenjun and Patel, Manooshree and Hartmann, Bjoern and Zamfirescu-Pereira, JD and Chasins, Sarah and Ranade, Gireeja},
  booktitle={The 5th Workshop on Mathematical Reasoning and AI at NeurIPS 2025}, 
  year={2025}
}

@article{jiang2023multilingual,
  title={Multilingual mathematical autoformalization},
  author={Jiang, Albert Q and Li, Wenda and Jamnik, Mateja},
  journal={arXiv preprint arXiv:2311.03755},
  year={2023}
}

@article{li2024autoformalize,
  title={Autoformalize mathematical statements by symbolic equivalence and semantic consistency},
  author={Li, Zenan and Wu, Yifan and Li, Zhaoyu and Wei, Xinming and Yang, Fan and Zhang, Xian and Ma, Xiaoxing},
  journal={Advances in Neural Information Processing Systems},
  volume={37},
  pages={53598--53625},
  year={2024}
}

@inproceedings{tarrach2024more,
  title={More details, please: Improving autoformalization with more detailed proofs},
  author={Tarrach, Guillem and Jiang, Albert Q and Raggi, Daniel and Li, Wenda and Jamnik, Mateja},
  booktitle={AI for Math Workshop@ ICML 2024},
  year={2024}
}

@article{ying2024lean,
  title={Lean workbook: A large-scale lean problem set formalized from natural language math problems},
  author={Ying, Huaiyuan and Wu, Zijian and Geng, Yihan and Wang, Jiayu and Lin, Dahua and Chen, Kai},
  journal={Advances in Neural Information Processing Systems},
  volume={37},
  pages={105848--105863},
  year={2024}
}

@article{gao2024herald,
  title={Herald: A natural language annotated Lean 4 dataset},
  author={Gao, Guoxiong and Wang, Yutong and Jiang, Jiedong and Gao, Qi and Qin, Zihan and Xu, Tianyi and Dong, Bin},
  journal={arXiv preprint arXiv:2410.10878},
  year={2024}
}

@article{wu2022autoformalization,
  title={Autoformalization with large language models},
  author={Wu, Yuhuai and Jiang, Albert Qiaochu and Li, Wenda and Rabe, Markus and Staats, Charles and Jamnik, Mateja and Szegedy, Christian},
  journal={Advances in Neural Information Processing Systems},
  volume={35},
  pages={32353--32368},
  year={2022}
}

@article{azerbayev2023proofnet,
  title={Proofnet: Autoformalizing and formally proving undergraduate-level mathematics},
  author={Azerbayev, Zhangir and Piotrowski, Bartosz and Schoelkopf, Hailey and Ayers, Edward W and Radev, Dragomir and Avigad, Jeremy},
  journal={arXiv preprint arXiv:2302.12433},
  year={2023}
}

@article{jiang2022draft,
  title={Draft, sketch, and prove: Guiding formal theorem provers with informal proofs},
  author={Jiang, Albert Q and Welleck, Sean and Zhou, Jin Peng and Li, Wenda and Liu, Jiacheng and Jamnik, Mateja and Lacroix, Timoth{\'e}e and Wu, Yuhuai and Lample, Guillaume},
  journal={arXiv preprint arXiv:2210.12283},
  year={2022}
}

@article{wang2024theoremllama,
  title={Theoremllama: Transforming general-purpose llms into lean4 experts},
  author={Wang, Ruida and Zhang, Jipeng and Jia, Yizhen and Pan, Rui and Diao, Shizhe and Pi, Renjie and Zhang, Tong},
  journal={arXiv preprint arXiv:2407.03203},
  year={2024}
}

@article{huang2024mustard,
  title={Mustard: Mastering uniform synthesis of theorem and proof data},
  author={Huang, Yinya and Lin, Xiaohan and Liu, Zhengying and Cao, Qingxing and Xin, Huajian and Wang, Haiming and Li, Zhenguo and Song, Linqi and Liang, Xiaodan},
  journal={arXiv preprint arXiv:2402.08957},
  year={2024}
}

@article{moore2025evaluating,
  title={Evaluating Autoformalization Robustness via Semantically Similar Paraphrasing},
  author={Moore, Hayden and Shah, Asfahan},
  journal={arXiv preprint arXiv:2511.12784},
  year={2025}
}

@misc{pramaanalabs2026,
  key          = {PramaanaLabs},
  title        = {Pramaana Labs},
  year         = {2026},
  howpublished = {\url{https://www.pramaanalabs.ai/}},
  note         = {Accessed: 2026-05-08}
}

@inproceedings{lu2025formalalign,
  title     = {{FormalAlign}: Automated Alignment Evaluation for Autoformalization},
  author    = {Lu, Jianqiao and Wan, Yingjia and Huang, Yinya and Xiong, Jing
               and Liu, Zhengying and Guo, Zhijiang},
  booktitle = {International Conference on Learning Representations (ICLR)},
  year      = {2025},
  eprint    = {2410.10135},
  archivePrefix = {arXiv}
}

@inproceedings{yang2023leandojo,
  title     = {{LeanDojo}: Theorem Proving with Retrieval-Augmented Language Models},
  author    = {Yang, Kaiyu and Swope, Aidan M and Gu, Alex and Chalamala, Rahul
               and Song, Peiyang and Yu, Shixing and Godil, Saad
               and Prenger, Ryan and Anandkumar, Anima},
  booktitle = {Advances in Neural Information Processing Systems},
  volume    = {36},
  year      = {2023}
}

@article{welleck2023llmstep,
  title={{LLMSTEP}: {LLM} proofstep suggestions in {Lean}},
  author={Welleck, Sean and Saha, Rahul},
  journal={arXiv preprint arXiv:2310.18457},
  year={2023}
}

@article{thoma2022lean,
  author  = {Thoma, Athina and Iannone, Paola},
  title   = {Learning about Proof with the Theorem Prover {LEAN}:
             The Abundant Numbers Task},
  journal = {International Journal of Research in Undergraduate
             Mathematics Education},
  volume  = {8},
  pages   = {64--93},
  year    = {2022},
  doi     = {10.1007/s40753-021-00140-1}
}

@article{wemmenhove2022waterproof,
  author  = {Wemmenhove, Jelle and Beurskens, Thijs and McCarren, Sean
             and Moraal, Jan and Tuin, David and Portegies, Jim},
  title   = {Waterproof: Educational Software for Learning How to
             Write Mathematical Proofs},
  journal = {arXiv preprint arXiv:2211.13513},
  year    = {2022}
}

@inproceedings{keenan2024hazel,
  author    = {Keenan, Matthew and Omar, Cyrus},
  title     = {Learner-Centered Design Criteria for Classroom
               Proof Assistants},
  booktitle = {Workshop on Human Aspects of Types and Reasoning
               Assistants ({HATRA})},
  year      = {2024},
  url       = {https://hazel.org/papers/learner_centered_proofs_hatra_2024.pdf}
}

@book{sieg1990proof,
  title={Proof theory},
  author={Sieg, Wilfried},
  year={1990},
  publisher={Carnegie Mellon University}
}

@article{xin2024deepseekprover,
  title={DeepSeek-Prover: Advancing Theorem Proving in LLMs through Large-Scale Synthetic Data},
  author={Xin, Huajian and Guo, Daya and Shao, Zhihong and Ren, Zhizhou and Zhu, Qihao and Liu, Bo and Ruan, Chong and Li, Wenda and Liang, Xiaodan},
  journal={arXiv preprint arXiv:2405.14333},
  year={2024}
}

@article{yang2024formal,
  title   = {Formal Mathematical Reasoning: A New Frontier in {AI}},
  author  = {Yang, Kaiyu and Poesia, Gabriel and He, Jingxuan and Li, Wenda and Lauter, Kristin and Chaudhuri, Swarat and Song, Dawn},
  journal = {arXiv preprint arXiv:2412.16075},
  year    = {2024},
  url     = {https://arxiv.org/abs/2412.16075}
}

@article{ospanov2025hermes,
  title   = {{HERMES}: Towards Efficient and Verifiable Mathematical Reasoning in {LLMs}},
  author  = {Ospanov, Azim and Feng, Zijin and Sun, Jiacheng and Bai, Haoli and Shen, Xin and Farnia, Farzan},
  journal = {arXiv preprint arXiv:2511.18760},
  year    = {2025},
  url     = {https://arxiv.org/abs/2511.18760}
}

@misc{hekmatnejad2019modelchecking,
  title         = {Model Checking Clinical Decision Support Systems Using {SMT}},
  author        = {Mohammad Hekmatnejad and Andrew M. Simms and Georgios Fainekos},
  year          = {2019},
  eprint        = {1901.04545},
  archivePrefix = {arXiv},
  primaryClass  = {cs.SE},
  doi           = {10.48550/arXiv.1901.04545},
  url           = {https://arxiv.org/abs/1901.04545},
  note          = {arXiv preprint arXiv:1901.04545 (v2, revised 4 March 2019)}
}

@misc{singh2026guarantees,
  title         = {Toward Guarantees for Clinical Reasoning in Vision Language
                   Models via Formal Verification},
  author        = {Singh, Vikash and Ganguly, Debargha and Yu, Haotian and
                   Zhou, Chengwei and Singh, Prerna and Lee, Brandon and
                   Chaudhary, Vipin and Datta, Gourav},
  year          = {2026},
  eprint        = {2602.24111},
  archivePrefix = {arXiv},
  primaryClass  = {cs.CV},
  doi           = {10.48550/arXiv.2602.24111},
  url           = {https://arxiv.org/abs/2602.24111}
}

\clearpage
\onecolumn

\appendix 
\section{Problem Statements}
\label{appendix:problems}

The following ten problems from EECS\,127 were used in our evaluation. Each problem was attempted by $n=10$ students.

\begin{enumerate}

\item \textbf{Q1} \\
Show that the following inequality hold for any vector \(\vec{x} \in \mathbb{R}^n \): \(\quad \frac{1}{\sqrt{n}} \|\vec{x}\|_2 \leq \|\vec{x}\|_\infty \)

\item \textbf{Q2} \\
Show that the following inequality hold for any vector \(\vec{x} \in \mathbb{R}^n \): \(\quad \|\vec{x}\|_\infty \leq \|\vec{x}\|_2 \)

\item \textbf{Q3}\\
Show that the following inequality hold for any vector \(\vec{x} \in \mathbb{R}^n \): \(\quad \|\vec{x}\|_2\leq \|\vec{x}\|_1 \)

\item \textbf{Q4} \\
Show that the following inequality hold for any vector \(\vec{x} \in \mathbb{R}^n \): \(\quad \|\vec{x}\|_1\leq \sqrt{n}\|\vec{x}\|_2 \)

\item \textbf{Q5}\\
Let \(A \in S^n\) (i.e., the set of \(n \times n\) symmetric matrices) and 
\((\lambda_1, \vec{v_1})\), \((\lambda_2, \vec{v_2})\) \(\lambda_1 \neq \lambda_2\) 
be distinct eigenpairs of \(A\) where \(\vec{v_1} \neq 0 \text{ and } \vec{v_2} \neq 0 \). 

Show that \(\vec{v_1}^\top \vec{v_2} = 0\) 
i.e., eigenspaces corresponding to distinct eigenvalues are mutually orthogonal. 
Note: this proof is part of the spectral theorem, so please do not use the spectral theorem in the proof.

\item \textbf{Q6} \\
Let $P$ be an $m \times n$ matrix. Show that the Frobenius norm satisfies the following property of a norm:
i. $ \left\lVert P \right\rVert_F = 0 $ if and only if $ P = 0 $

\item \textbf{Q7}  \\
Let $P$ be an $m \times n$ matrix. Show that the Frobenius norm satisfies the following property of a norm:
ii. For all $a \in \mathbb{R}$, we have $\left\lVert aP \right\rVert_F = |a| \left\lVert P \right\rVert_F$

\item \textbf{Q8} \\
Let $P$ and $Q$ be $m \times n$ matrices. Show that the Frobenius norm satisfies the following property of a norm:
iii. $\left\lVert P + Q \right\rVert_F \le \left\lVert P \right\rVert_F + \left\lVert Q \right\rVert_F$

\item \textbf{Q9} \\
Let $P$ be an $m \times n$ matrix. Write $\left\lVert P \right\rVert_F^2$ in terms of singular values $\sigma_1,\ldots,\sigma_n$. 

HINT: The cyclic property of traces might be helpful: $\text{tr}(PQR) = \text{tr}(RPQ) = \text{tr}(QRP)$.

\item \textbf{Q10} \\
In this problem, we will analyze properties of positive semidefinite (PSD) matrices. A symmetric matrix \( M \in \mathbb{R}^{n \times n}\) is a PSD matrix if \(\vec{x}^{\top} M \vec{x} \geq 0 \quad \text{for all } \vec{x} \in \mathbb{R}^n,\) and we denote that as \(M \succeq 0 \quad \text{or} \quad M \in \mathbb{S}_+^n.\)
Assume \(A \in \mathbb{R}^{n \times n}\) is a symmetric matrix.
Show that if \(A \succeq 0\) then all diagonal entries of \(A\) are non-negative, \(A_{ii} \geq 0.\)

\item \textbf{Q11} \\
Show that for any matrix \(Q \in \mathbb{R}^{m \times n}\), if \( A = Q^{\top}Q \) then \( A \in \mathbb{S}_+^n.\)
Note: If we take \(Q = P \in \mathbb{S}_+^n\), we have \( A = P^2 \in \mathbb{S}_+^n;\) this proves the other direction of the above ``if-and-only-if.''

\item \textbf{Q12} \\
Let \(A \in \mathbb{R}^{n \times n}\) be a symmetric matrix with eigenvalue-eigenvector pairs given by \((\lambda_0, \vec{v}_0), \ldots, (\lambda_{n-1}, \vec{v}_{n-1})\), where \(\lambda_0 > \cdots > \lambda_{n-1}\). Consider the problem \[p^* = \max_{\vec{x} \in \mathbb{R}^n} \vec{x}^\top A \vec{x} \quad \text{s.t. } \quad\|\vec{x}\|_2^2 = 1, \quad \vec{x}^\top \vec{v}_0 = 0.\quad\]  Show that \(p^* = \lambda_1\)

HINT: First find an \(\vec{x}\) which is feasible and \(\vec{x}^\top A \vec{x} = \lambda_1\). Then show that \(p^* \leq \lambda_1\).

\item \textbf{Q13} \\
Let \(A \in \mathbb{R}^{n \times n}\) be a symmetric matrix with eigenvalue-eigenvector pairs given by \((\lambda_0, \vec{v}_0), \ldots, (\lambda_{n-1}, \vec{v}_{n-1})\), where \(\lambda_0 > \cdots > \lambda_{n-1}\). Consider the problem \[p^* = \max_{\vec{x} \in \mathbb{R}^n} \vec{x}^\top A \vec{x} \quad \text{s.t. } \quad\|\vec{x}\|_2^2 = 1, \quad \vec{x}^\top \vec{v}_0 = 0.\quad\] Firstly, show that \(p^* \geq \lambda_1\) HINT: First find an \(\vec{x}\) which is feasible and \(\vec{x}^\top A \vec{x} = \lambda_1\).

\item \textbf{Q14} \\
Let \(A \in \mathbb{R}^{n \times n}\) be a symmetric matrix with eigenvalue-eigenvector pairs given by \((\lambda_0, \vec{v}_0), \ldots, (\lambda_{n-1}, \vec{v}_{n-1})\), where \(\lambda_0 > \cdots > \lambda_{n-1}\). Consider the problem \[p^* = \max_{\vec{x} \in \mathbb{R}^n} \vec{x}^\top A \vec{x} \quad \text{s.t. } \quad\|\vec{x}\|_2^2 = 1, \quad \vec{x}^\top \vec{v}_0 = 0.\quad\] Then, show that \(p^* \leq \lambda_1\)

\end{enumerate}

Noted that Q13 and Q14 are two sub parts of Q12, as in the classroom deployment we found the Q12 takes too long time to complete in the study, therefore we cut them into 2. As we are not measuring the whole problem completeness, thus we did not discard the study data for Q12.

\newpage
\section{More interface screenshots}
\label{appendix_more_examples}
\begin{figure}[h]
    \centering
    \includegraphics[width=0.9\linewidth]{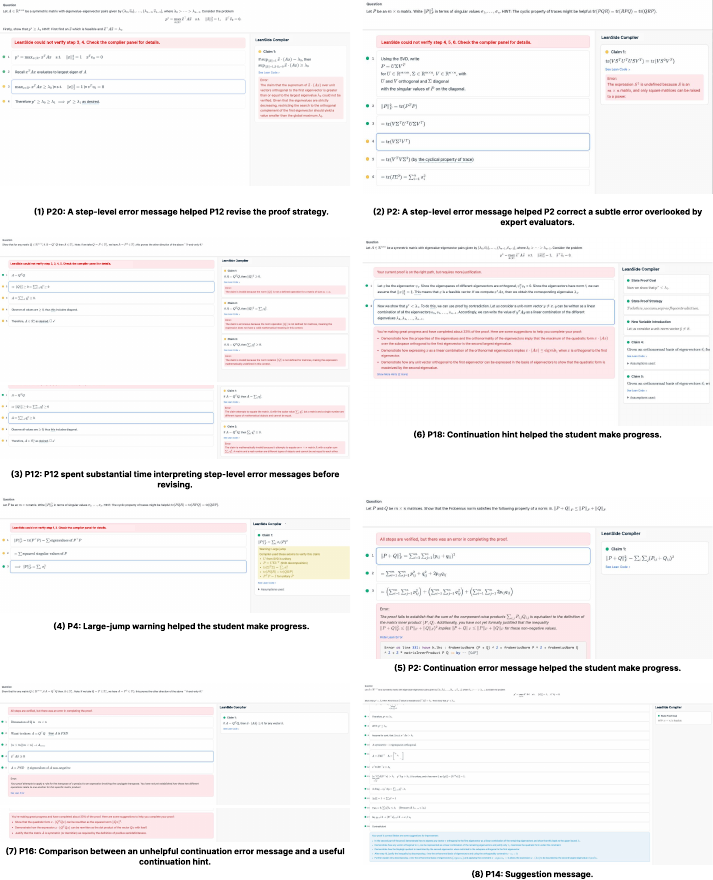}
    \caption{examples}
    \label{fig:placeholder}
\end{figure}
\newpage
\section{Participants' demographic information}
\label{appendix_participants}
\begin{table}[htbp]
\centering
\small
\caption{Participant background and familiarity with Lean}
\label{tab:participant-background}
\begin{tabularx}{\linewidth}{
    @{}
    l
    >{\raggedright\arraybackslash}X
    c
    >{\raggedright\arraybackslash}X
    @{}
}
\toprule
Participant & Major & Year & Lean familiarity \\
\midrule
P2  & CS+math                         & 1      & Experienced with Lean introduction game \\
P3  & Aerospace Engineering \& EECS  & 4      & No \\
P4  & Chemistry                       & 3      & Tried another formal language \\
P6  & CS                              & 2      & Knew Lean before, but never tried to write in Lean \\
P7  & CS+math                         & 2      & Experienced with Lean introduction game \\
P8  & CS                              & 2      & Knew Lean before, but never tried to write in Lean \\
P9  & CS+math                         & 2      & No \\
P11 & CS                              & 2      & No \\
P12 & CS                              & 1      & Knew Lean before, but never tried to write in Lean \\
P13 & CS + data science               & 4      & No \\
P14 & EECS                            & Master & No \\
P15 & CS+datascience                  & 1      & No \\
P16 & Data science                    & 2      & No \\
P18 & CS \& statistic                 & 2      & No \\
P19 & Transportation Engineering      & PhD    & No \\
P20 & Data science                    & 2      & No \\
P21 & Data science                    & 2      & No \\
P22 & CS+math                         & 1      & Knew Lean before, but never tried to write in Lean \\
P23 & CS+math                         & 4      & Knew Lean before, but never tried to write in Lean \\
P24 & CS+math                         & 4      & No \\
P25 & Bio engineering                 & 3      & No \\
P26 & CS + business                   & 2      & Knew Lean before, but never tried to write in Lean \\
P27 & Business                        & 2      & No \\
\bottomrule
\end{tabularx}
\end{table}

Exam grade distribution:
\begin{figure}[b]
    \centering
    \includegraphics[width=0.6\linewidth]{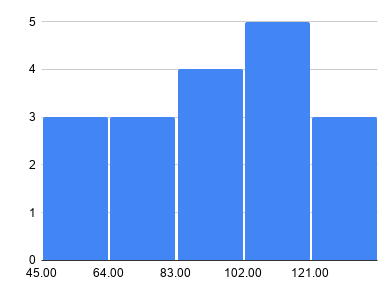}
    \caption{This figure shows student's grade distribution. }
    \label{fig:grade}
\end{figure}

\end{document}